\documentclass{aa}  
\usepackage[colorlinks, allcolors=blue]{hyperref}
\usepackage{amsmath} 
\usepackage{graphicx}
\usepackage{xcolor}
\usepackage{txfonts}
\usepackage{longtable}
\usepackage{booktabs}
\usepackage{geometry}
\usepackage{datetime}
\usepackage{lscape}
\usepackage{float}
\usepackage{placeins}
\usepackage{amssymb}
\usepackage[normalem]{ulem}

\begin{document} 

   \title{A deep X-ray and UV look into the reflaring stage of the accreting millisecond pulsar SAX J1808.4$-$3658}

\author{C. Ballocco \inst{1,2}
\and A. Papitto\inst{1}
\and A. Miraval Zanon \inst{3}
\and G. Illiano \inst{4}
\and T. Di Salvo \inst{5}
\and F. Ambrosino \inst{1}
\and L. Burderi \inst{6,9}
\and S. Campana \inst{4}
\and F. Coti Zelati \inst{7,8,4}
\and A. Di Marco \inst{11}
\and C. Malacaria \inst{1}
\and M. Pilia \inst{10}
\and J. Poutanen \inst{12}
\and T. Salmi \inst{13}
\and A. Sanna \inst{6}
}

\institute{INAF - Osservatorio Astronomico di Roma, Via Frascati 33, I-00078 Monte Porzio Catone (RM), Italy\\
\email{caterina.ballocco@inaf.it} 
\and Dipartimento di Fisica, Sapienza Università di Roma, Piazzale Aldo Moro 5, I-00185 Rome, Italy 
\and ASI - Agenzia Spaziale Italiana, Via del Politecnico snc, I-00133 Rome, Italy 
\and INAF - Osservatorio Astronomico di Brera, Via Bianchi 46, I-23807, Merate (LC), Italy
\and Dipartimento di Fisica e Chimica - Emilio Segrè, Università di Palermo, Via Archirafi 36, 90123 Palermo, Italy
\and Dipartimento di Fisica, Università degli Studi di Cagliari, SP Monserrato-Sestu km 0.7, I-09042 Monserrato, Italy
\and Institute of Space Sciences (ICE, CSIC), Campus UAB, Carrer de Can Magrans s/n, E-08193 Barcelona, Spain
\and Institut d’Estudis Espacials de Catalunya (IEEC), E-08860 Castelldefels (Barcelona), Spain
\and INAF/IASF Palermo, Via Ugo La Malfa 153, 90146 Palermo, Italy
\and INAF - Osservatorio Astronomico di Cagliari, Via della Scienza 5, 09047 Selargius (CA), Italy
\and INAF Istituto di Astrofisica e Planetologia Spaziali, Via del Fosso del Cavaliere 100, 00133 Roma, Italy
\and Department of Physics and Astronomy, FI-20014 University of Turku, Finland
\and Department of Physics, P.O. Box 64, FI-00014 University of Helsinki, Finland
}
\date{}

\authorrunning{Ballocco et al.}
\titlerunning{A deep X-ray and UV look into the reflaring stage of the AMSP SAX J1808.4$-$3658} 

\abstract 
{
We present a detailed X-ray and UV high-time-resolution monitoring of the final reflaring phase of the 2022 outburst of the accreting millisecond pulsar SAX~J1808.4$-$3658, based on simultaneous \textit{XMM-Newton} and \textit{Hubble Space Telescope} (\textit{HST}) observations.
The uninterrupted coverage provided by \textit{XMM-Newton} enabled a detailed characterization of the spectral and temporal evolution of the source X-ray emission, as the flux varied by approximately one order of magnitude.
We detected coherent X-ray pulsations during the whole X-ray observation, down to a 0.5--10~keV luminosity of $L_{\text{X(low)}\,0.5-10} \simeq 6.21^{+0.20}_{-0.15} \times 10^{34} \, d^2_{3.5}\,\mathrm{erg \, s^{-1}}$; this is among the lowest ever observed in this source during the outburst state. 
At the lowest flux levels, we observed significant variations in pulse amplitude and phase. These variations were anticorrelated with the X-ray source flux. We found a sharp phase jump of $\sim 0.4$ cycles, accompanied by a doubling of the pulse amplitude and a softening of the X-ray emission. We interpreted changes in the X-ray pulse profiles as drifts of emission regions on the neutron-star surface, driven by an increase in the inner-disk radius when the mass-accretion rate decreased.
The dependence of the pulse phase on the X-ray flux was consistent with a magnetospheric radius scaling as $R_{\rm{m}} \propto \dot{M}^{\Lambda}$, with $\Lambda = -0.17(9)$, which is in broad agreement with theoretical predictions.
Simultaneous \textit{HST} observations confirmed the presence of significant UV pulsations at an X-ray luminosity approximately a factor of two lower than during the 2019 outburst, extending the range of mass accretion rates at which UV pulsations have been detected.
The measured pulsed UV luminosity, $L_{\text{pulsed}}^{\text{UV}} = 1.1(3) \times 10^{32} \, \text{erg} \, \text{s}^{-1}$, was consistent with that observed during the 2019 outburst. Yet, such a UV luminosity exceeds the predictions of standard emission models, as further confirmed by the shape of the pulsed spectral energy distribution. }

   
   \maketitle

\section{Introduction}
Accreting millisecond pulsars (AMSPs) are weakly magnetized ($B\sim 10^{8-9}\,$G) neutron stars (NSs) that accrete matter from a low-mass companion ($\lesssim 1\,\text{M}_\odot$) via an accretion disk (e.g., \citealt{PatrunoWatts_2021, DiSalvoSanna_2022, CampanaDiSalvo}). Their high spin frequencies result from a previous billion-year-long evolutionary phase of mass accretion \citep{Alpar_1982, BHATTACHARYA_1991}.
In this scenario, known as ``recycling,'' slowly rotating radio pulsars are spun up to millisecond periods through the accretion of matter and angular momentum transferred from the companion star via Roche-lobe overflow. AMSPs are all transient systems that spend most of their life in quiescence. Occasionally, they start several-week-long mass-accretion outbursts, during which the X-ray luminosity increases by a few orders of magnitude up to $\simeq 10^{36-37}\,\mathrm{erg\,s^{-1}}$. During these outbursts, the pulsar magnetosphere channels the accreting matter towards the NS magnetic poles, producing coherent X-ray pulsations at a frequency higher than $\sim 100\, \text{Hz}$.
More than two dozen AMSPs have been discovered so far \citep[see, e.g., the most recent discovery in][]{Ng_2024}.

The radius at which the NS magnetosphere starts controlling the dynamics of the infalling matter, $R_{\rm{in}}$, decreases with the mass-accretion rate \citep{Gosh_1979,Spruit_1993}. Toward the end of an X-ray outburst, $R_{\rm{in}}$ expands and might exceed the corotation radius, $R_{\rm{co}}$. When this happens, a centrifugal barrier is expected to hinder mass accretion onto the NS, in the so-called propeller regime \citep{Illarionov_1975, Stella_1994, Campana_1998}. However, \citet{Spruit_1993} showed that accretion may proceed at a reduced rate even when $R_{\rm{in}}\gtrsim R_{\rm{co}}$, because the centrifugal barrier is too weak to expel matter from the system.
In this regime, a “trapped” or “dead” disk may form \citep{Dangelo_2010, D'angelo_2012}, where matter accumulates due to inefficient angular-momentum loss, allowing only a small fraction to reach the NS surface. Magnetohydrodynamic simulations by \citet{Romanova_2018} showed that either a weak or strong propeller state may be realized. In the former ($R_{\rm in}\gtrsim  R_{\rm co}$), part of the inflowing material can penetrate the centrifugal barrier and accrete along the magnetic-field lines. In the strong propeller regime ($R_{\rm in}\gg R_{\rm co}$), most of the matter is expelled from the system, and accretion onto the stellar surface is largely suppressed.

As they fade into quiescence, X-ray observations of AMSPs are a powerful tool with which to test these scenarios. SAX~J1808.4$-$3658 (hereafter SAX~J1808) is the best studied AMSP during the 11 approximately one-month-long outbursts shown from its discovery in 1998 \citep{wijnands_1998}. After reaching a peak luminosity of $\sim10^{36-37} \,\mathrm{erg\,s^{-1}}$ a few days after the outburst onset, this source typically shows a slow decay ($\sim$ 10 days) followed by a rapid drop (3--5 days) (see, e.g., Fig.~\ref{fig:lctot}). Before returning to a quiescent level of $\sim 5 \times 10^{31}\,\mathrm{erg\, s^{-1}}$ \citep{Campana_2004, Stella_2000}, SAX~J1808 displays peculiar luminosity variations dubbed “reflares” for a month or longer. During reflares, the X-ray luminosity varies in the range of $\sim 10^{32-35} \,\mathrm{erg\,s^{-1}}$ on timescales of approximately one-to-two days \citep{patruno_2009b, patruno_2016,PatrunoWatts_2021}. It has been proposed that reflares originate from small changes in the disk density at the end of an outburst that increase the outer-disk temperature, partially ionizing hydrogen \citep{patruno_2009b, patruno_2016}. This leads to a rapid increase in accretion into the inner disk and a subsequent temporary rebrightening. 
The detection of X-ray pulsations, the aperiodic variability and the stable spectral distribution of the X-ray emission observed during the reflares of SAX~J1808 suggest that the innermost regions of the accretion flow remain close to the corotation radius even at the lowest X-ray luminosity attained (a few $\times\,10^{34} \,\mathrm{erg\,s^{-1}}$; \citealt{patruno_2009b,patruno_2016}). 
Coherent X-ray pulsations have also been observed at similarly low luminosities of a few $\times 10^{33}\, \mathrm{erg\,s^{-1}}$ in two additional AMSPs \citep{Bult_2019b, Illiano_inprep}. Altogether, these observed properties suggest that channeled accretion in AMSPs can persist down to very low mass-accretion rates at which propeller ejection of matter would be rather expected.

SAX~J1808 was also the first AMSP to show coherent pulsations at optical and UV wavelengths during the rising and final stages of its 2019 outburst, respectively \citep{Ambrosino_2021}. The observed highly pulsed optical and UV luminosities ($L_{\rm pulsed}^{\rm opt}\approx3\times10^{31}$\, erg\, s$^{-1}$, $L_{\rm pulsed}^{\rm UV}\approx2\times10^{32}$\, erg\, s$^{-1}$) exceeded the values expected in the case of cyclotron emission from accretion columns by roughly $50-100$ times, and the expected thermal output at those wavelengths by several orders of magnitude.
Optical and UV pulsations have also been detected from the transitional millisecond pulsar PSR~J1023+0038 \citep{Ambrosino_2017, Papitto_2019, Jaodand_2021, Ambrosino_2021, Miraval_2022}, and from the redback PSR~J2339-0533 \citep{Papitto_2025}. In the former case, the high optically pulsed luminosities observed required a nonstandard interpretation in terms of synchrotron emission from the interface between the relativistic wind of a rotation-powered pulsar and the accretion flow \citep{Papitto_2019, Veledina_2019, Baglio2023, Baglio_2024}. However, this scenario assumes that the NS is active, as a rotation-powered pulsar and cannot apply to SAX~J1808, where simultaneous X-ray pulsations indicate that the accretion flow extends down to the NS surface. The origin of the pulsed optical and UV emission in SAX~J1808 thus remains uncertain.

In this paper, we report on a 125\, ks-long \textit{XMM-Newton} observation and a simultaneous 2.2~ks \textit{Hubble Space Telescope} (\textit{HST}) observation, performed during the reflaring stage of the outburst observed in SAX~J1808 in 2022 \citep{Illiano_2023}. \textit{XMM-Newton}'s uninterrupted coverage enabled a high-resolution study of the spectral and temporal evolution of the source X-ray emission as its flux varies by roughly one order of magnitude down to a few $10^{34} \mathrm{erg\,s^{-1}}$. The simultaneous \textit{HST} observation allowed us to extend the range of mass-accretion rates at which UV pulsations have been studied by a factor of approximately two and to perform the first simultaneous study in X-ray and UV bands.

\section{Observations}
\subsection{XMM-Newton}
\begin{figure}
\centering
\resizebox{\hsize}{!}{\includegraphics[clip=true]
{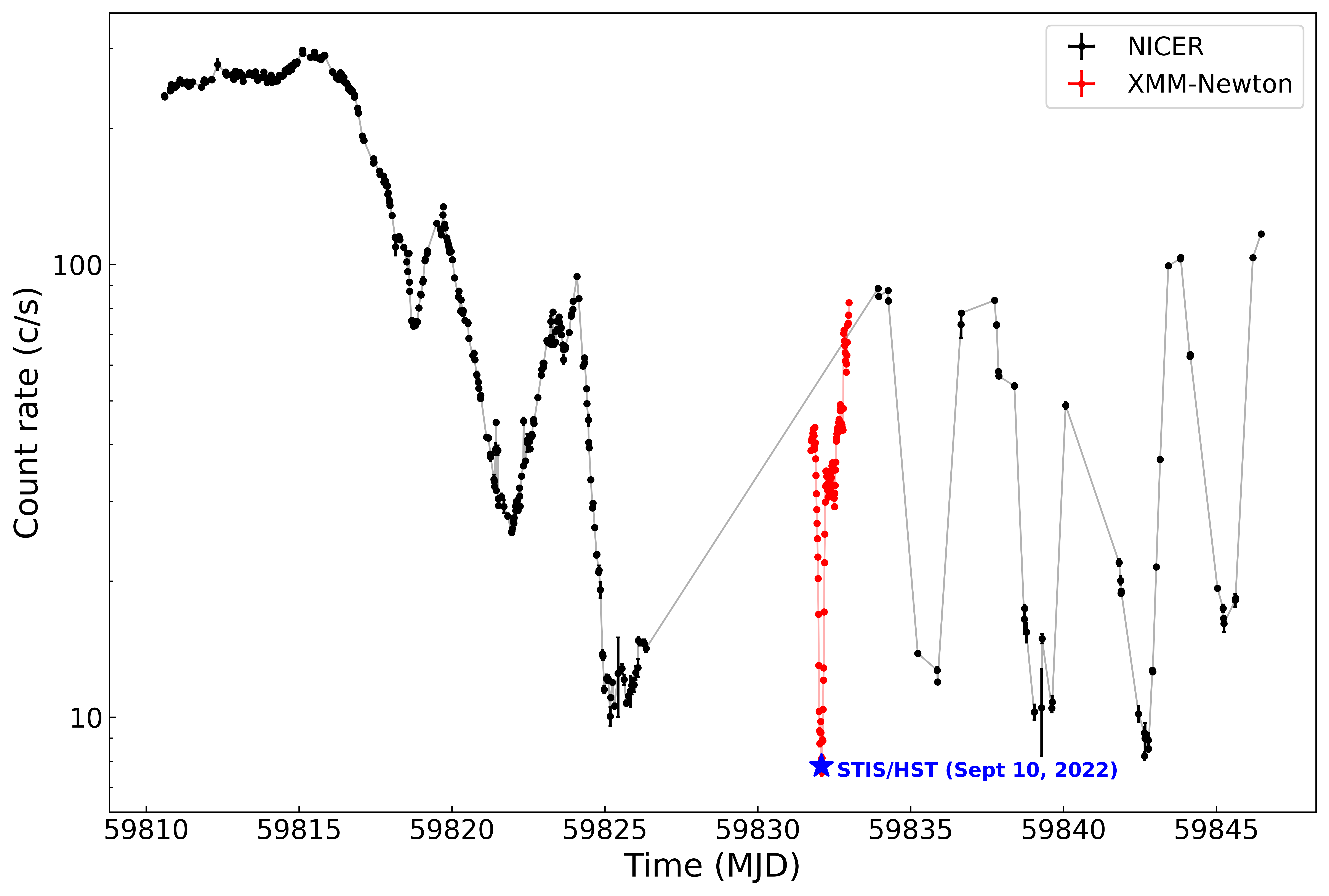}}
\caption{
\footnotesize
Light curve of 2022 outburst of SAX~J1808 in 0.5--10~keV energy band, using $\mathrm{1\,ks}$ bins. \textit{NICER} observations are shown in black, and \textit{XMM-Newton} observations are given in red. The blue star marks the epoch of the \textit{HST}/STIS observation (2022 September 10). We rescaled the \textit{XMM-Newton} count rate to the \textit{NICER} count rate using a conversion factor of 1.44, which was calculated with the WebPIMMS tool\protect\footnotemark\, and a power-law model with a photon index of $\Gamma = 2.04$ (Table~\ref{tab:spectra}).} 
\label{fig:lctot}
\end{figure}

\begin{figure}
\centering
\resizebox{\hsize}{!}{\includegraphics[clip=true]{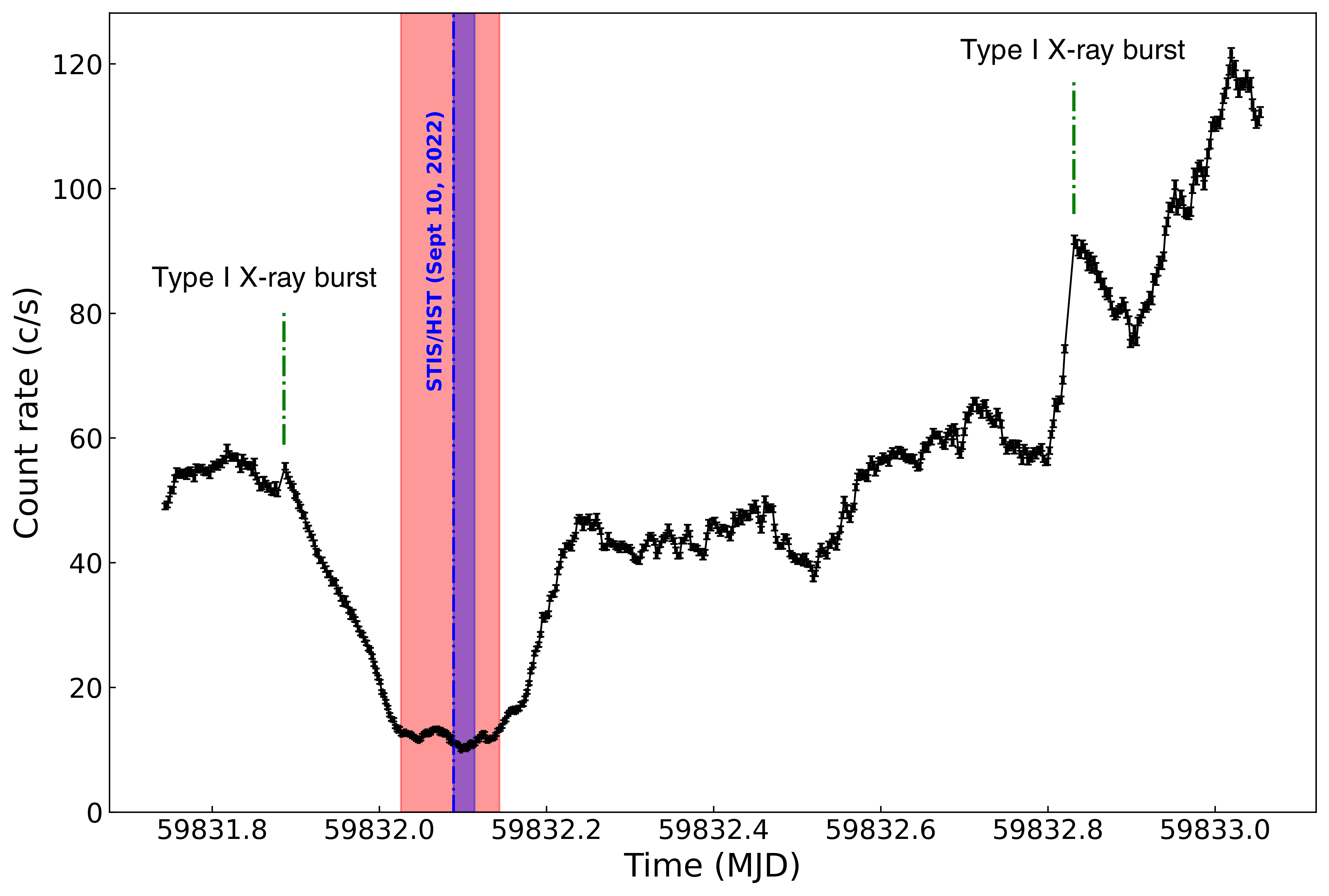}}
\caption{
\footnotesize
Background-subtracted 0.5--10~keV \textit{XMM-Newton}/EPIC-pn light curve of 2022 outburst of SAX~J1808 measured using $200\,\text{s}$-long bins. A red band indicates the low-flux interval, and a blue band shows the interval of the \textit{HST} observation discussed in this paper (see Sect.~\ref{sec:UV} for details). Two Type I X-ray bursts, detected at 59831.87 and 59832.82 MJD, are not plotted. Green dash-dotted lines indicate the epochs of their occurrence.
} 
\label{fig:Xlc}
\end{figure}

\textit{XMM-Newton} \citep{Jansen} observed SAX~J1808 for 125~ks starting on 2022 September 9 at 14:22:09 (UTC) (Obs.ID. 0884700801, PI: A. Papitto), during the final stage of the outburst (see red points in Fig.~\ref{fig:lctot}). Figure \ref{fig:Xlc} shows the background-subtracted 0.5--10 $\mathrm{keV}$ EPIC-pn light curve. 
We used the Science Analysis System (SAS; v20.0.0) to process and reduce the data. We barycentered the photon arrival times observed by \textit{XMM-Newton} using the \texttt{barycen} tool, with the JPL DE-405 Solar System ephemerides and the source position derived by \citet{Bult_2020}.
The EPIC-pn camera operated in timing mode to achieve the 29.5\,$\mu$s temporal resolution needed to study source variability and minimize photon pile-up, and it was equipped with a thick filter.
The effective exposure was reduced to 108.2~ks after excluding soft-proton flaring episodes, identified when the EPIC-pn 10--12~keV count rate exceeded $0.8\, \mathrm{c\,s^{-1}}$. 
In timing mode, spatial information along one of the optical axis was discarded to allow a faster readout. As a result, the source appears on the EPIC-pn detector as a bright strip rather than its usual circular shape. The maximum number of counts were recorded in pixels characterized by RAWX coordinates 37 and 38.
EPIC-pn events were extracted from a 21-pixel-wide region (1 pixel $\simeq 4.1 ''$) around the source position, spanning from $\mathrm{RAWX =28}$ to 48. We estimated the background far from the source, in a three-pixel-wide region centered on $\mathrm{RAWX = 4}$. 

During the \textit{XMM-Newton} pointing, SAX~J1808 exhibited two Type~I X-ray bursts at MJD 59831.87 and 59832.82, which will be discussed in a dedicated upcoming publication. After obtaining the light curve, we created good time intervals (GTIs), discarding time intervals starting $\mathrm{10\,s}$ before and ending $\mathrm{150\,s}$ after each burst onset to perform spectral and timing analyses of the non-bursting emission. 
Following the recommended \textit{XMM-Newton}/SAS procedures\footnote{\url{https://www.cosmos.esa.int/web/xmm-newton/sas-threads}}, we extracted the source and background spectra from the cleaned event file, considering only single- and double-pixel events (PATTERN $\leq$ 4) and filtering out spurious events (FLAG = 0). We generated spectral response files using the tasks \texttt{rmfgen} and \texttt{arfgen}. The EPIC-pn spectra were rebinned using the SAS \texttt{specgroup} tool to have at least 25 background-subtracted counts per spectral channel and to sample the instrument's resolution full width at half maximum with no more than three channels at any energy.
\footnotetext{\url{https://heasarc.gsfc.nasa.gov/cgi-bin/Tools/w3pimms/w3pimms.pl}.}

\subsection{HST/STIS}\label{sec:hst}
The Space Telescope Imaging Spectrograph (STIS) on board \textit{HST} observed SAX~J1808 for 2.2 ks on 2022 September 10, starting at 02:05:41 (UTC) (GO/DD-17245, PI Miraval Zanon). The UV observation occurred during the interval of minimum X-ray flux, as shown by the blue band in Fig.~\ref{fig:Xlc}.
We removed an $\sim 400$ s bump in the recorded flux at MJD 59832.103, which occurred when the observatory grazed the region of the South Atlantic Anomaly (\textit{HST} help desk, private communication).  The resulting UV light curve of the source observed with STIS is consistent with a constant count rate of $\sim 45\,\mathrm{c\,s^{-1}}$.
The spectroscopic observation was performed in TIME-TAG mode, achieving a time resolution of 125 $\mu \text{s}$. The G230L grating, equipped with a $52 \times 0.2\,\text{arcsec}^2$ slit, provided a spectral resolution of $\sim$\,500 over the nominal (first-order) range.
The total photon count rate was $R_{\rm{{HST}}} = (46.62 \pm 0.14)\,\mathrm{c\, s^{-1}}$, with approximately 35\% attributed to background radiation ($\mathrm{BKG_{HST} = (16.66 \pm 0.09)\, c\,s^{-1}}$).

We extracted the source photons across 19 slit channels ($990-1010$) and in the $165-310$~nm wavelength range. This selection isolates the source signal, minimizes background contributions, and avoids noise arising from the G230L grating's reduced response at the edges of the wavelength range.
We estimated the background signal by selecting photons in the STIS slit channels outside the source region ($200-900$ and $1100-2000$) in order to have high statistics and avoid the source contribution. We then averaged the resulting value and normalized it to the number of source slit channels.
To analyze \textit{HST}/STIS data, we first corrected the position of the slit channels using the Python custom-built external function \texttt{stis\_photons}\footnote{\url{https://github.com/Alymantara/stis_photons}}. 
UV photon arrival times were then corrected to the Solar System barycenter (SSB) using the \texttt{ODELAYTIME} task (subroutine available in the IRAF/STDAS software package) and JPL DE200 ephemerides. Note that the difference between the JPL DE200 and DE405 ephemerides is negligible compared to the $\mathrm{\sim 1\,s}$ uncertainty on the absolute timing of the \textit{HST} data (\textit{HST} help desk, private communication).

\subsection{NICER}
We analyzed the \textit{Neutron star Interior Composition Explorer} \citep[\textit{NICER};][]{Gendreau_2012} observations obtained during the 2019 and 2022 outbursts and previously presented by \citet{Bult_2020} and \citet{Illiano_2023}. In 2019, \textit{NICER} observed SAX~J1808 from July 30 (MJD 58695) until November 8 (MJD 58795; ObsIDs 205026 and 258401).
In 2022, we observed the source between August 19 (MJD 59810) and October 31 (MJD 59883; ObsIs 505026 and 557401). We processed and corrected the data according to the procedures described by \citet{Bult_2020} and \citet{Illiano_2023}.
All data were processed using the standard \textit{NICER} analysis software \texttt{NICERDAS}, part of HEASoft (v6.35.2). We processed the data using the \texttt{nicerl2} task, selecting events in the $0.5-10$\,keV energy range. Photon arrival times for both datasets were corrected to the Solar System barycenter using the JPL DE405 ephemerides and the \texttt{BARYCORR} tool. We then adopted the source coordinates from \citet{Hartman_2008} and \citet{Bult_2020}, respectively, for the 2019 and 2022 \textit{NICER} datasets, to allow a direct comparison with the analyses presented by \citet{Bult_2020} and \citet{Illiano_2023}.

\section{Coherent timing analysis} \label{sec:Xrays}
\subsection{X-ray timing analysis}
\begin{figure}
\centering
\resizebox{\hsize}{!}{\includegraphics[clip=true]{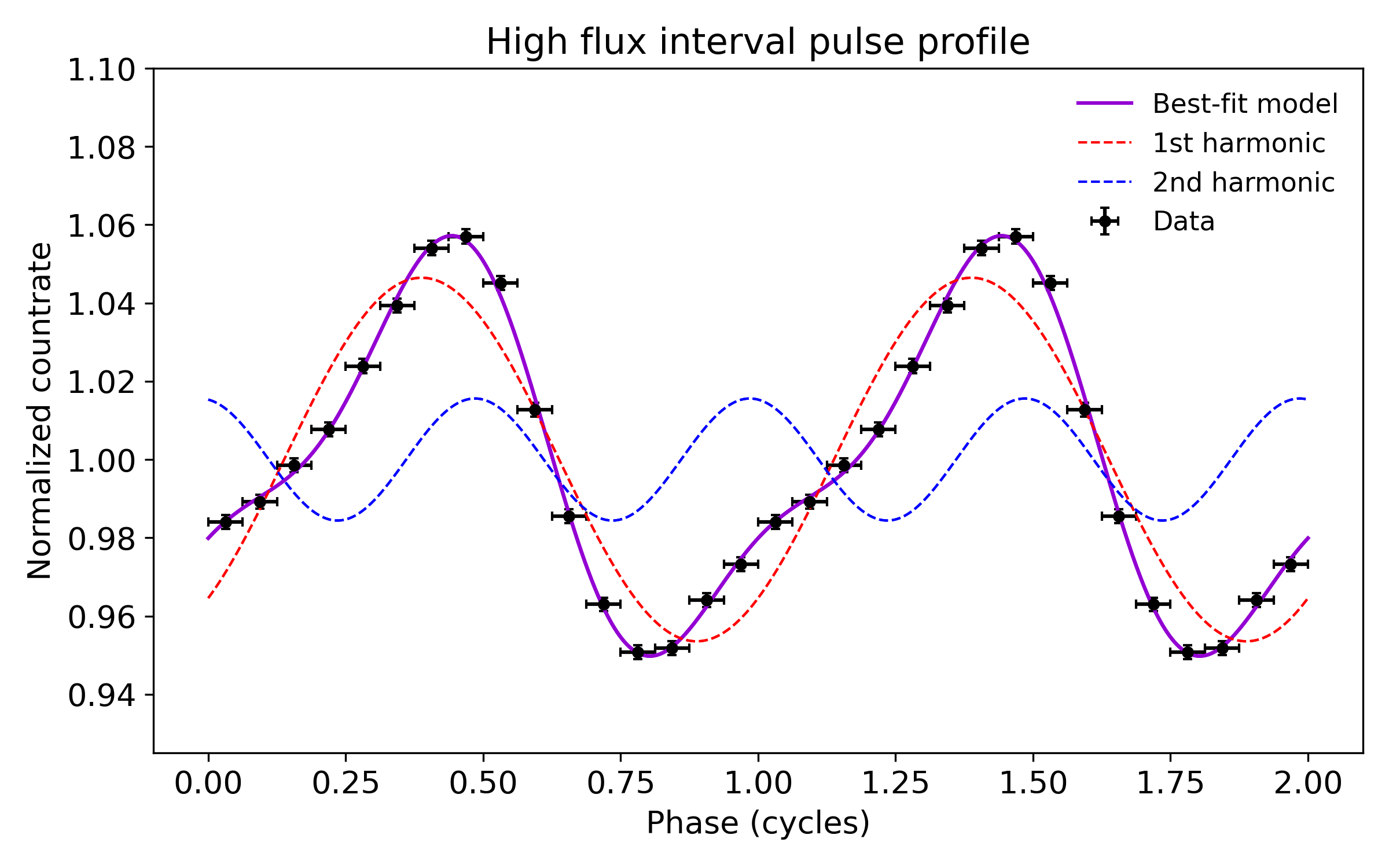}}
\hspace{5mm}
\resizebox{\hsize}{!}{\includegraphics[clip=true]{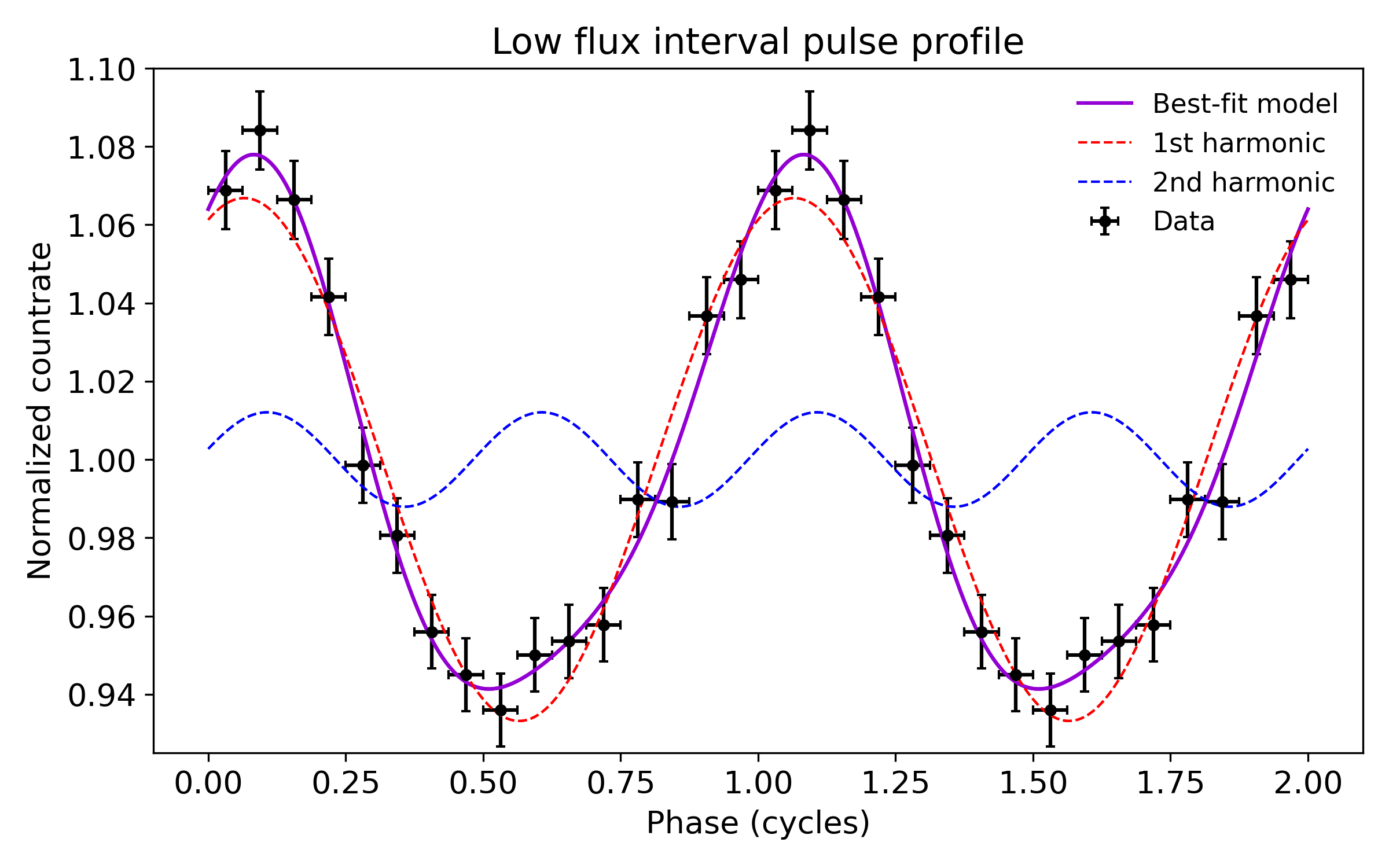}}
\caption{X-ray pulse profiles obtained by folding the high- and low-flux intervals of \textit{XMM-Newton} observation into 16 phase bins at a spin frequency of $\nu_F = 400.975209557\,\text{Hz}$. The solid line represents the best-fit model (constant plus two harmonics), while the dashed red and blue lines indicate the first and second harmonic contributions, respectively. For clarity, two cycles are shown.}
\label{fig:Xpulseprofiles}
\end{figure}

\begin{figure}
\resizebox{\hsize}{!}{\includegraphics[clip=true]{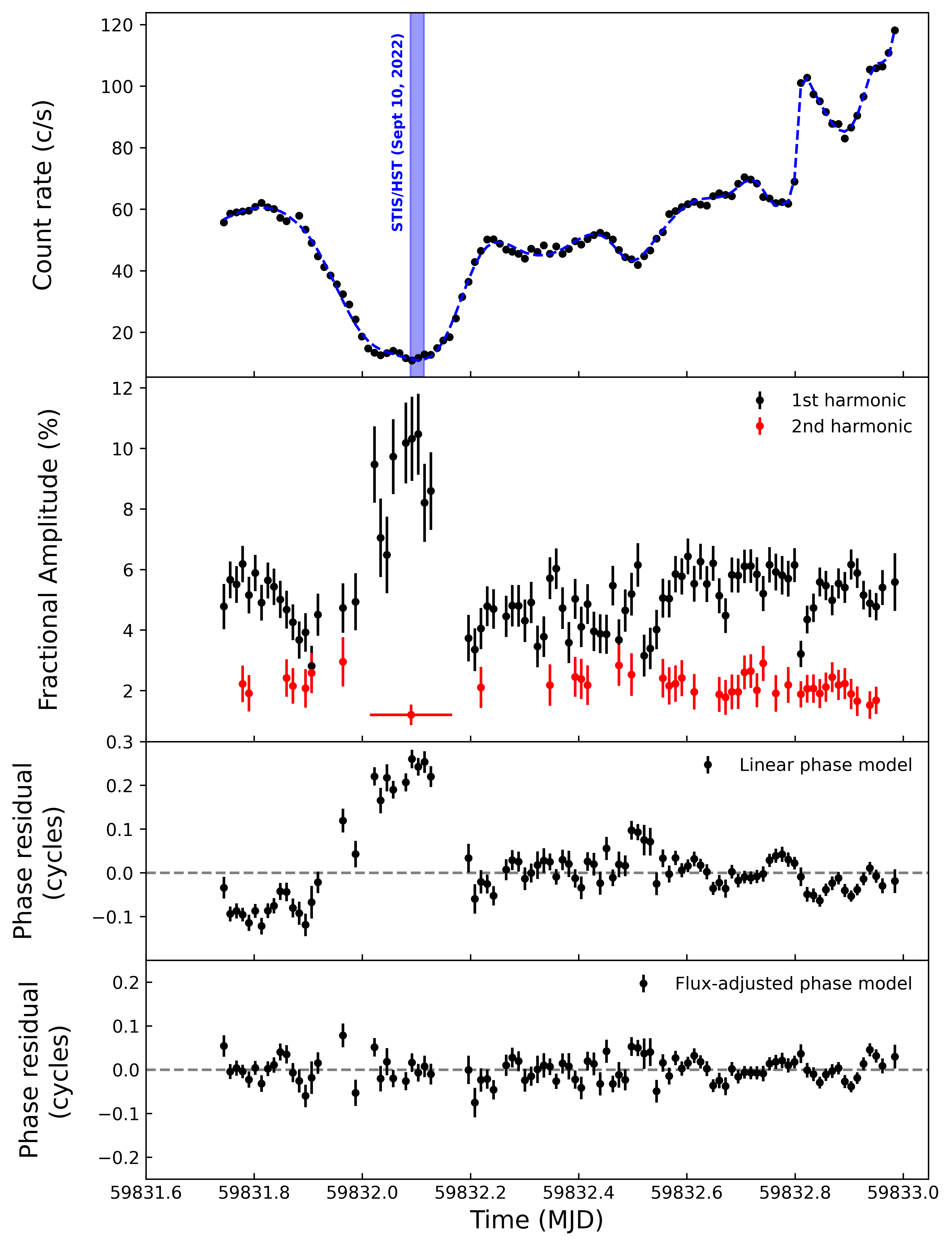}}
\caption{
\footnotesize
Top panel: 0.5--10~keV \textit{XMM-Newton}/EPIC-pn light curve using 1\,ks bins, with Type I X-ray bursts removed. The dashed blue line shows the spline interpolation obtained with the \texttt{UnivariateSpline} function from \texttt{scipy.interpolate}. The blue band indicates the interval of the \textit{HST} observation discussed in this paper (see Sect.~\ref{sec:UV} for details).
Second panel: Pulse fractional amplitude for the first harmonic (black dots) and the second harmonic (red dots). 
Third and fourth panels: Phase residuals relative to a linear model and a model that includes a phase-flux correlation term, respectively. Note that, where not visible, the error bars are smaller than the data points.
}
\label{fig:residuals}
\end{figure}

We performed a coherent timing analysis of the 2022 outburst of SAX~J1808 using the \textit{XMM-Newton}/EPIC-pn high-time-resolution dataset. We first corrected the photon arrival times for the Doppler delays introduced by the binary motion using the orbital parameters measured by \citet{Illiano_2023} from the pulse-timing analysis of 2022 \textit{NICER} data.
We created pulse profiles by epoch folding $1000\,\text{s}$ time intervals in 16 phase bins around the best available estimate of the spin frequency ($\nu_F=400.975209557\,\text{Hz}$ from \citealt{Illiano_2023}). 
We adopted a bin length of 1000\,s to ensure adequate statistics to detect the pulsation and track its amplitude and phase variations (see Fig.~\ref{fig:residuals}). Two harmonic components at most are usually required to satisfactorily model the pulse profile of SAX~J1808 during such short intervals \citep[see, e.g.,][]{Hartman_2008}, justifying the choice of sampling it in 16 phase bins.
Only statistically significant pulse profiles were selected, namely folded profiles with a ratio between the sinusoidal amplitude and the corresponding 1$\sigma$ error equal to or greater than three.
The second panel of Fig.~\ref{fig:residuals} shows the fractional amplitudes of the first and second harmonic. We focused the analysis of this work on the former, as it is significantly stronger.
During the minimum of the X-ray flux (MJD 59832.02 to 59832.14; see the red band in Fig.~\ref{fig:Xlc}), the amplitude of the fundamental (black dots) roughly doubled up to a value of $\sim10$\%.
Figure \ref{fig:Xpulseprofiles} shows the X-ray pulse profiles obtained by epoch by folding the high and low-flux intervals (red band in Fig.~\ref{fig:Xlc}) of the \textit{XMM-Newton} observation separately.

\begin{figure}
\resizebox{\hsize}{!}{\includegraphics[clip=true]{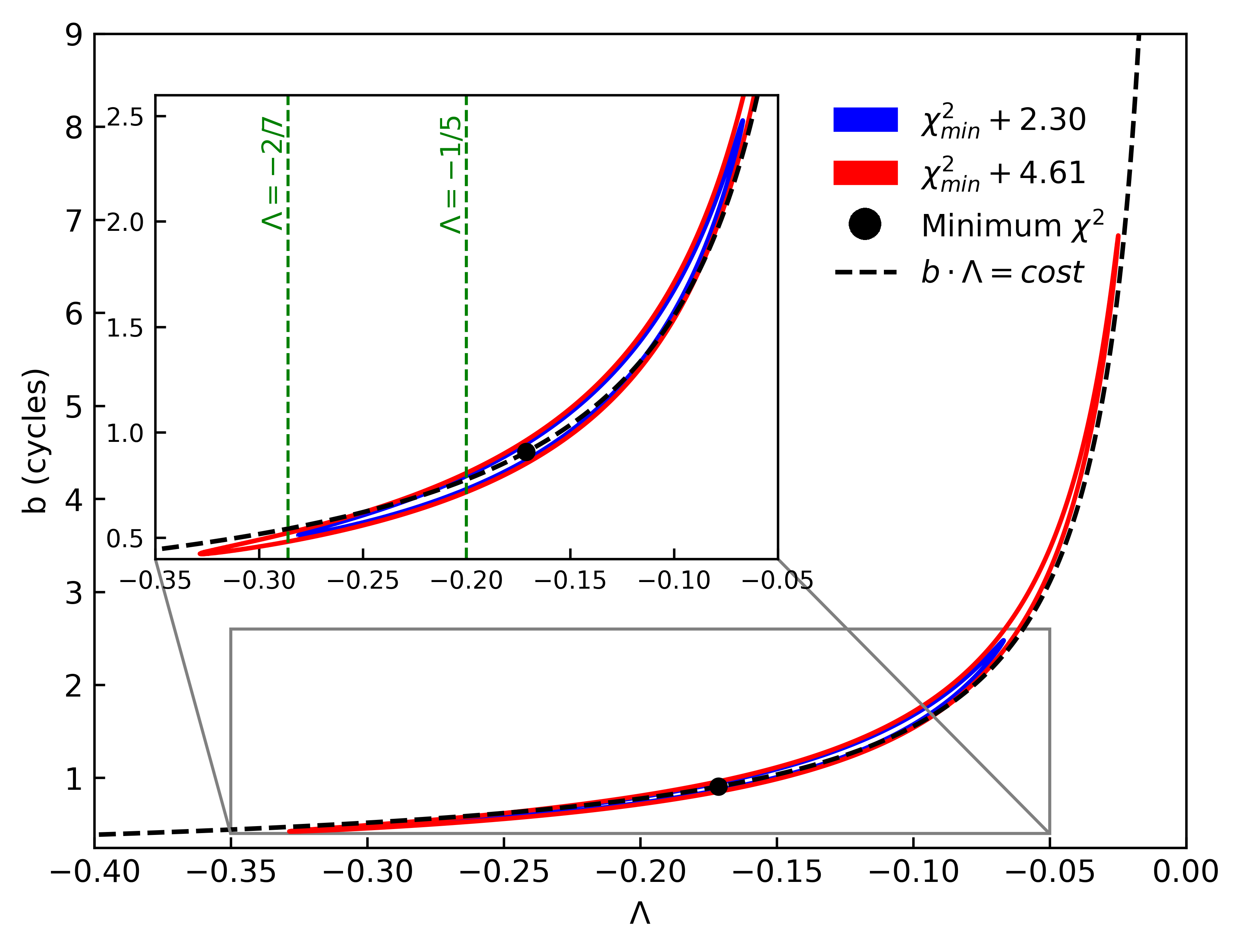}}
\caption{
\footnotesize
Contour levels of $\chi^2$ obtained by varying both parameters $b$ and $\Lambda$ in the flux-adjusted phase fit. The contour levels are shown for $\chi^2_{\min} + 2.3$ (blue) and $\chi^2_{\min} + 4.61$ (red), corresponding to confidence levels of 68\% and 90\%, respectively, for a fit with two parameters \citep{Lampton_1976}. The black dot indicates the values of $b$ and $\Lambda$ corresponding to the minimum $\chi^2$, while the dashed black line represents the contour corresponding to the constant value given by the product $b \times \Lambda$.}
\label{fig:chi2curve}
\end{figure}

We first modeled the evolution of the pulse phase measured over the first harmonic using the following relation \citep[e.g.,][]{Burderi_2007, Sanna_2022b, Papitto_2007}:
\begin{equation}
\phi(t)=\phi_0-\Delta\nu(t-T_0)-\frac{1}{2} \dot{\nu}(t-T_0)^2+R_{\rm{orb}}(t).
    \label{eq:fit}
\end{equation}
Here, $T_0$ is the reference epoch for the timing solution, $\phi_0$ is the pulse phase at $T_0$, $\Delta\nu=\nu(T_0)-\nu_F$ represents the difference between the frequency at the reference epoch and the spin frequency used to epoch-fold the data, and $\dot{\nu}$ is the average spin-frequency derivative. The term $R_{\rm{orb}}$ models the phase residuals arising from differences between the adopted orbital parameters and the actual ones \citep[see, e.g.,][]{deeter}.
We first considered a constant frequency model ($\dot{\nu} = 0$), obtaining a value of $\nu(T_0)$ that was not compatible with that expected according to the timing solution reached by \citet{Illiano_2023} (see Table~\ref{tab:orbpar}).
\begin{table}
\centering
\caption{Timing solution for SAX~J1808 during 2022 outburst.}
\begin{tabular}{lc}
\toprule
Parameter & Value \\
\midrule
Epoch, $T_0$ [MJD]  & 59831.7381191 \\
$P_{\rm{orb}}$ [s] & 7249.19(44) \\
$a \sin{i}/c$ [lt-s]  & 0.06283(11) \\
$T_{\rm{asc}}$ [MJD] & 59810.6178(13) \\
\toprule
Linear phase model \\
\midrule
$\nu$ [Hz]  & 400.97521088(95) \\
$\chi^2$/d.o.f  & 1419.2/92 \\
\\
\toprule
Flux-adjusted phase model \\
\midrule
$\nu$ [Hz] & 400.97520957(12) \\
$b$ & $0.9_{-0.4}^{+1.6}$ \\
$\Lambda$ & $-$0.17(9) \\
$\chi^2$/d.o.f & 164.45/90 \\
\bottomrule
\\
\end{tabular}
\tablefoot{To take into account the large value of the reduced $\chi^2$ obtained from the fit, we rescaled the uncertainties of the fit parameters by the square root of that value \citep[see, e.g.,][]{Finger_1999}. Uncertainties are the 1$\sigma$ statistical errors.
}
\label{tab:orbpar}
\end{table} 
Adding a spin frequency derivative term to the model reduced the residuals ($\Delta\chi^2=339$ for one degree of freedom less), but the modeling remained statistically unsatisfactory ($\chi^2_r=11.86$ over 91 degrees of freedom).
The residuals with respect to either a linear or a quadratic model show large deviations (see the third panel of Fig.~\ref{fig:residuals}). This phenomenon often affects the evolution of the pulse phase of AMSPs and is commonly referred to as timing noise \citep{Burderi_2006, Papitto_2007, Hartman_2008, Hartman_2009, Patruno_2009}. Timing noise is characterized by phase variations with amplitudes of up to several tenths of a cycle, affecting the measurement of the pulse frequency and its derivative.
Here, a large phase jump of $\sim 0.4$ cycles occurring around 59832 MJD, on a timescale of a few hours, was the most evident feature (see Fig.~\ref{fig:residuals}). This jump coincides with a decrease in the X-ray flux of the source by a factor of approximately three. When the flux returned to the initial values, the pulse phase also reverted to the estimates obtained before the jump.
To account for the phase shifts, we added a term describing a correlation between the X-ray flux and the pulse phase to Eq.\, \eqref{eq:fit}, $R_{\rm{flux}}(t)$, which is possibly related to hot spot drifts \citep{Patruno_2009}:
\begin{equation}
\phi(t)=\phi_0-\Delta\nu(t-T_0)+R_{\rm{orb}}(t)+R_{\rm{flux}}(t).
    \label{eq:fluxadjmod}
\end{equation}
Following \citet{Bult_2020}, we used $R_{\rm{flux}}(t)=b\,(F_{\rm{X}}/F_0)^\Lambda$, with $F_{\rm{X}}$ being the X-ray bolometric flux, approximated by the 0.5--10~keV count rate for simplicity, and $F_0$ the initial flux value at $t=T_0$. 
To apply such a model, we interpolated the light-curve data using the \texttt{UnivariateSpline} function (available in Python's \texttt{scipy.interpolate} library), as done by \citet{Illiano_2023}.
This method allows the flux values to be interpolated at the exact time points where phase data are available.
The estimates of the X-ray flux so obtained (plotted with a dashed blue line in the top panel of Fig.~\ref{fig:residuals}) were used in Eq.\,\eqref{eq:fluxadjmod} to fit the pulse phases, obtaining the residuals shown in the bottom panel of Fig.~\ref{fig:residuals}.
By adding this flux-phase correlation term, we obtained a reduced $\chi^2$ of 1.82 for 90 degrees of freedom (d.o.f.), showing a highly significant improvement over the quadratic phase model ($\Delta \chi^2=915$ for the addition of one free parameter), with a probability of $\sim 1.5\times10^{-38}$ of it being due to chance, according to an F-test. The best-fitting parameters of the flux-phase correlation term are $b = 0.9^{+1.6}_{-0.4}$ and $\Lambda = -0.17(9)$.
Even though the $\chi^2$ value obtained is still formally statistically unacceptable, the phase residuals no longer showed the phase jump, nor did they indicate the presence of significant structures or trends (Fig.~\ref{fig:residuals}, bottom panel).
Unlike the linear and quadratic models, the value of the spin frequency obtained from the flux-adjusted model (Table~\ref{tab:orbpar}) was consistent within the errors with that expected according to the timing solution reported in \citet{Illiano_2023}, thus supporting the validity of the corrected model.\\
To estimate the uncertainties on the parameters $b$ and $\Lambda$, we considered the contour levels of the fit $\chi^2$, $\chi^2_{\text{min}} + 2.3,$ and $\chi^2_{\text{min}} + 4.61$ (Fig.~\ref{fig:chi2curve}), corresponding to confidence levels of 68\% and 90\%, respectively, for a fit with two free parameters of interest \citep{Lampton_1976, Avni_1976, Yaqoob_1998}.
The shape of the contour plots strongly suggests a correlation $b \times \Lambda \simeq \rm{const}$ between the fit parameters.

For comparison, we applied the same coherent timing technique to investigate the phase evolution of \textit{NICER} data collected during the 2019 and 2022 outbursts. 
To avoid introducing significant uncertainties into the spline function, we filtered out data from the end of the outburst when the temporal gaps between consecutive observations exceeded three days. 
In both datasets, a flux-adjusted phase model again provided a better description of the observed phases than a linear or quadratic modeling.
We obtained $b = 0.75^{+0.30}_{-0.14}$ and $\Lambda = -0.14(5)$ for the 2019 dataset, and $b = 0.08^{+0.07}_{-0.03}$ and $\Lambda = -0.37(14)$ for the 2022 observation.
The values of $\Lambda$ are consistent, within uncertainties, with the value derived from the \textit{XMM-Newton} timing analysis. These results are consistent with those of \citet{Bult_2020}, where $\Lambda$ was fixed at $-1/5$. In contrast, the discrepancy with the values reported by \citet{Illiano_2023} arises from the different time interval considered for the phase fitting.

\begin{figure}
\resizebox{\hsize}{!}{\includegraphics[clip=true]{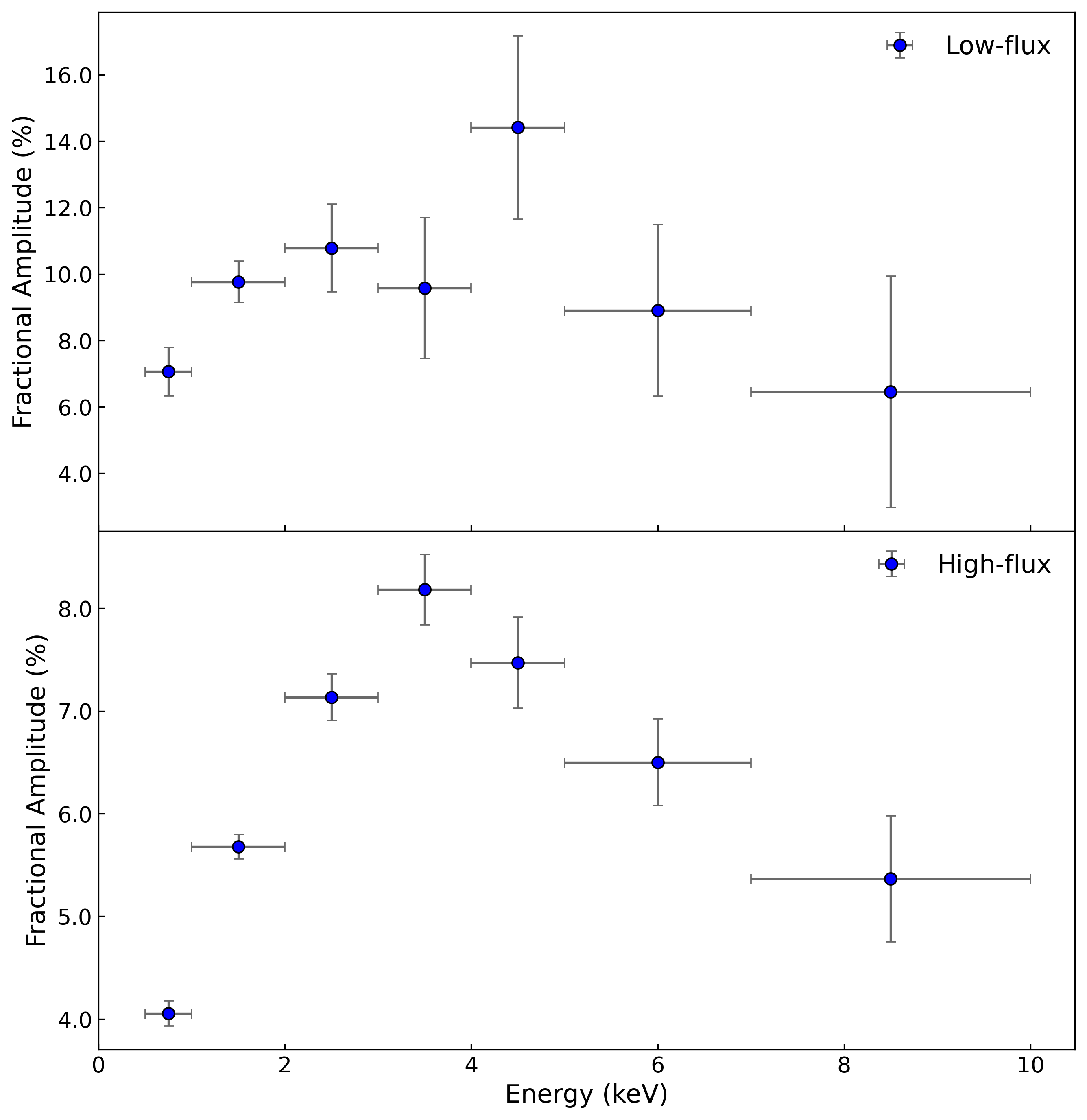}}
\caption{
\footnotesize
Fractional amplitude of fundamental harmonic of X-ray pulse profiles in seven energy bands (0.5--1, 1--2, 2--3, 3--4, 4--5, 5--7, 7--10~keV) for the low-flux part of the light curve (top panel) and for the high-flux emission (bottom panel).
}
\label{fig:ampvsenergy}
\end{figure}

Figure~\ref{fig:ampvsenergy} shows the fractional pulse amplitudes in seven energy bands, computed in the low-flux phase of the \textit{XMM-Newton} light curve (top panel), and in the high-flux phase of the X-ray emission (bottom panel).
The morphology we observed closely matches interval~4 of \citet{Bult_2020} (see their Fig.~2), corresponding to the reflare phase of the 2019 outburst, which shows the highest amplitudes and a fractional amplitude that increases with energy before returning to near-initial values at $\mathrm{\sim 10 \,keV}$. The amplitudes we observed are generally higher than those reported by \citet{Bult_2020}, although the relative increase in pulsed fraction from $\mathrm{\sim 1\,keV}$ to 2--3~keV is consistent across the two outbursts, both showing an increase of approximately 40\% relative to the initial values.

To compare the properties of the X-ray and UV pulsations, we measured the X-ray pulse amplitude in an interval strictly simultaneous to the \textit{HST} observations (marked by a blue band in Fig.~\ref{fig:residuals}). In that interval, we obtained $A_{\rm{X}}=(9.9\pm0.8)$\%. The 0.5--10~keV flux measured in that time interval using a spectral model described in Table~\ref{tab:spectra} was $F_{0.5-10}=(4.24^{+0.14}_{-0.10})\times 10^{-11}\, \mathrm{erg\,cm^{-2}\,s^{-1}}$. The pulsed 0.5--10~keV X-ray luminosity was then evaluated as $L_{\rm{pulsed}}^{\rm X} =(A_{\rm{X}}/\sqrt{2}) (4\pi d^2 F_{0.5-10}) = (4.4\pm0.4)\times 10^{33} \,d_{3.5}^2\, \text{erg s}^{-1}$, where $d_{3.5}$ is the distance to the source in units of 3.5 kpc \citep{galloway_2006}.

\subsection{UV timing analysis} \label{sec:UV}
\begin{figure}
\resizebox{\hsize}{!}{\includegraphics[clip=true]{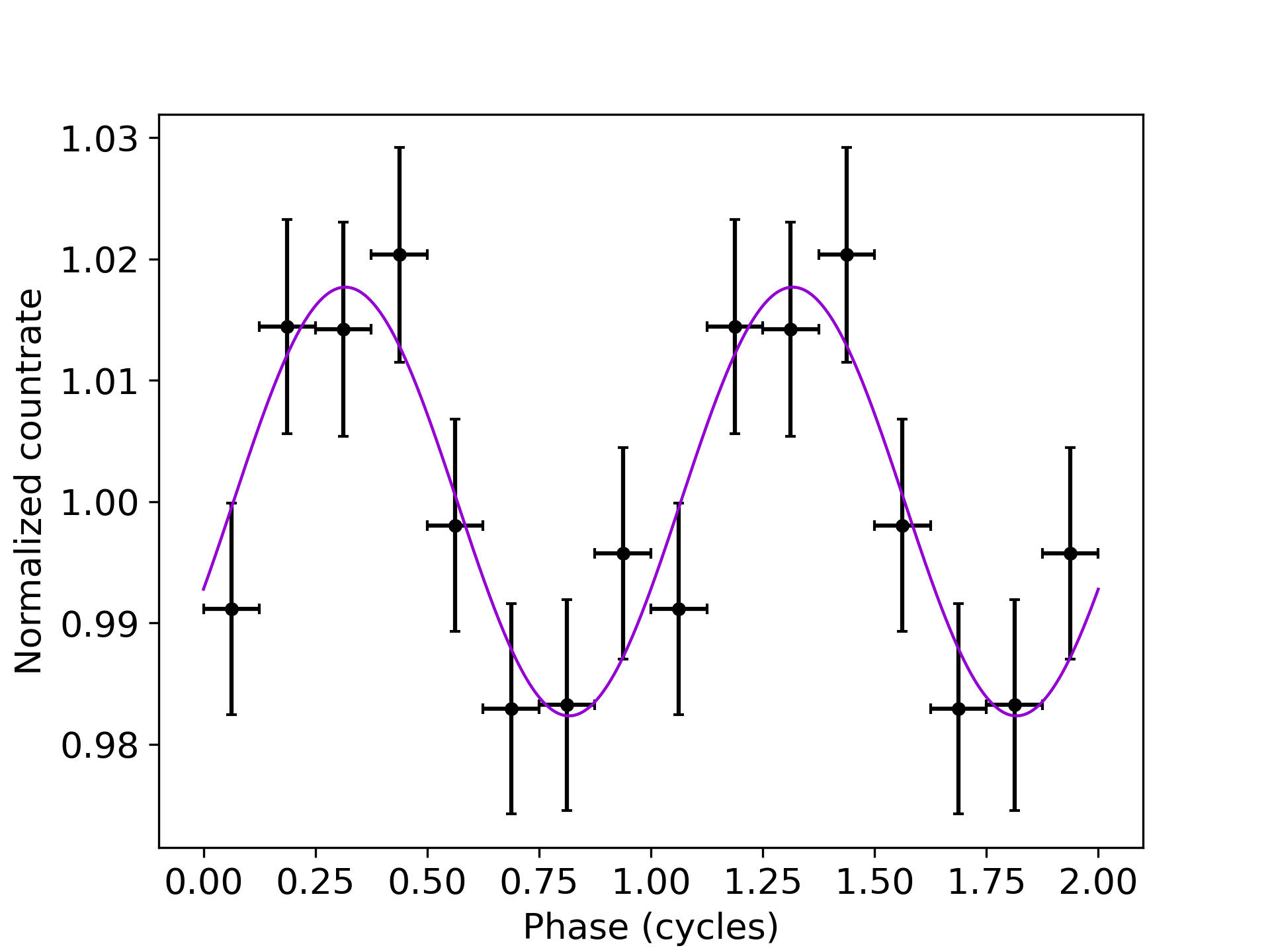}}
\caption{
\footnotesize
Pulse profile obtained by folding the whole \textit{HST}/STIS observation (GO/DD-17245, PI Miraval Zanon) into eight phase bins at a spin frequency of $\nu_F = 400.975209557\,\text{Hz}$. The solid line represents the best-fit model, consisting of a single sinusoidal component. For clarity, two cycles are shown.}
\label{fig:pulsprofUV}
\end{figure}

To measure the spin period observed during the $2.2$\,ks-long \textit{HST}/STIS observation, we performed an epoch folding search using $m=8$ phase bins and a period resolution of $\delta P_{\text{HST,EFS}} = P^2/2mT_{obs} = 1.7 \times 10^{-10}\,\text{s}$, where $\mathrm{T_{obs}}$ is the total exposure time of the \textit{HST} observation.
We thus obtained a best-fitting period of $P_{\text{HST,EFS}} = 2.4939194(6)\times10^{-3} \, \text{s}$. 
Following \citet{Leahy_1987}, the $1\sigma$ uncertainty reported was calculated using the following equation: $\mathrm{\sigma_P=(P^2/2\,T_{obs})}\cdot0.71(\chi^2_{\rm max}/m-1)^{-0.63}$, where the maximum of the $\chi^2$ value was found to be $\chi^2_{\rm max}\simeq 26.6$.
The period inferred from UV data is consistent, within uncertainties, with that obtained from X-ray data.
We then folded the \textit{HST} data at the best spin period. A single sinusoidal component describes the resulting UV pulsed profiles well, and an additional harmonic is not statistically justified, as it does not significantly improve the fit chi-square (see Fig.~\ref{fig:pulsprofUV}).
To take the background into account, we evaluated a correction factor as $N_{\rm{\gamma(tot)}}/(N_{\rm \gamma(tot)} - N_{\rm \gamma(bkg)})\simeq 1.556(5)$, where $N_{\rm{\gamma(tot)}}$ is the total number of detected photons and $N_{\rm{\gamma(bkg)}}$ the number of background photons.
After applying this correction, the root-mean-square (r.m.s.) amplitude resulted as $A^{\text{rms}}_{\text{UV}} = (1.9\pm 0.5)\%$.

\section{X-ray spectral analysis} \label{sec:spectra}
\begin{figure}
\centering
\resizebox{0.99\hsize}{!}{\includegraphics[clip=true]{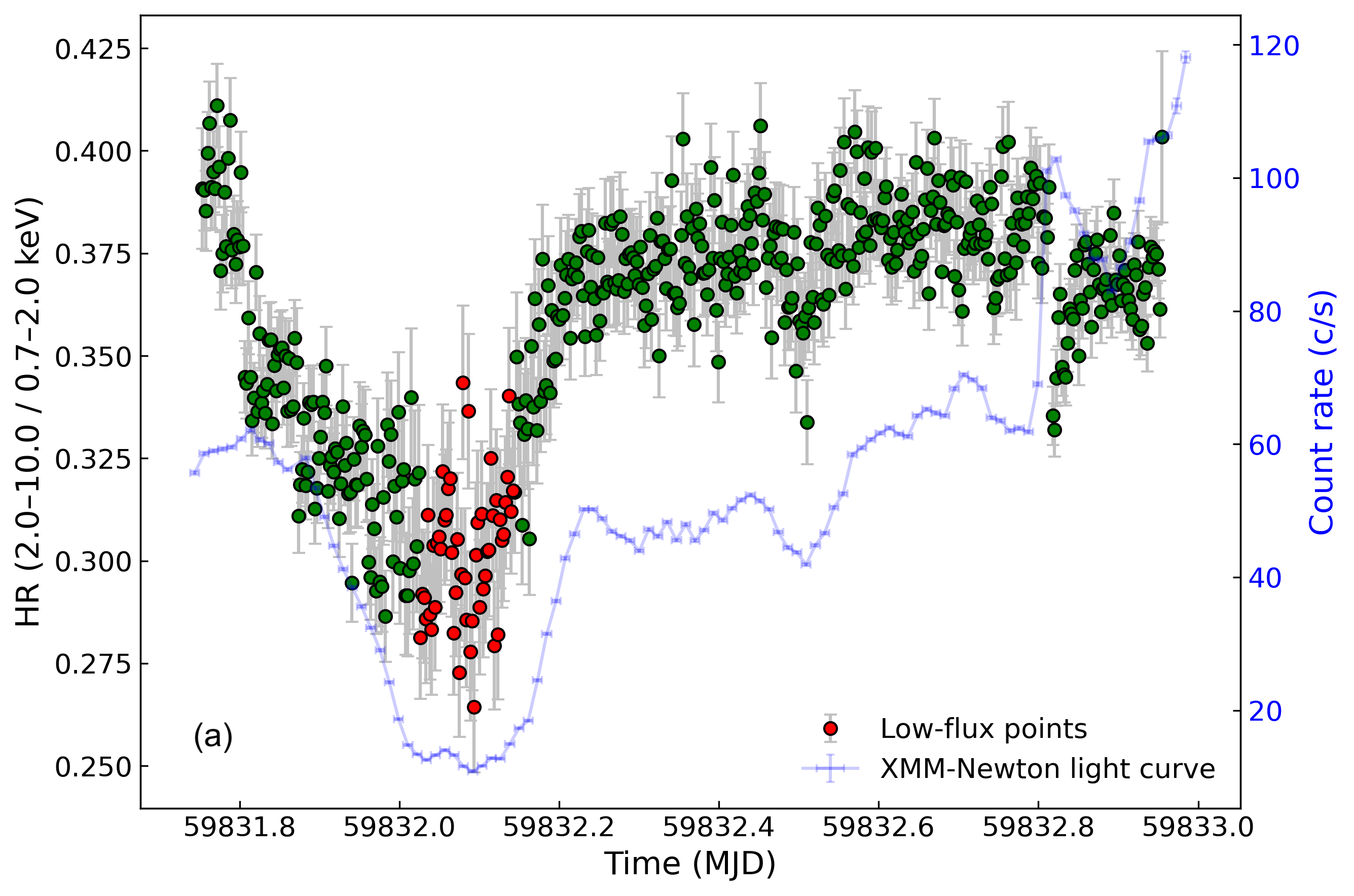}}
\hspace{5mm}
\resizebox{0.95\hsize}{!}{\includegraphics[clip=true]{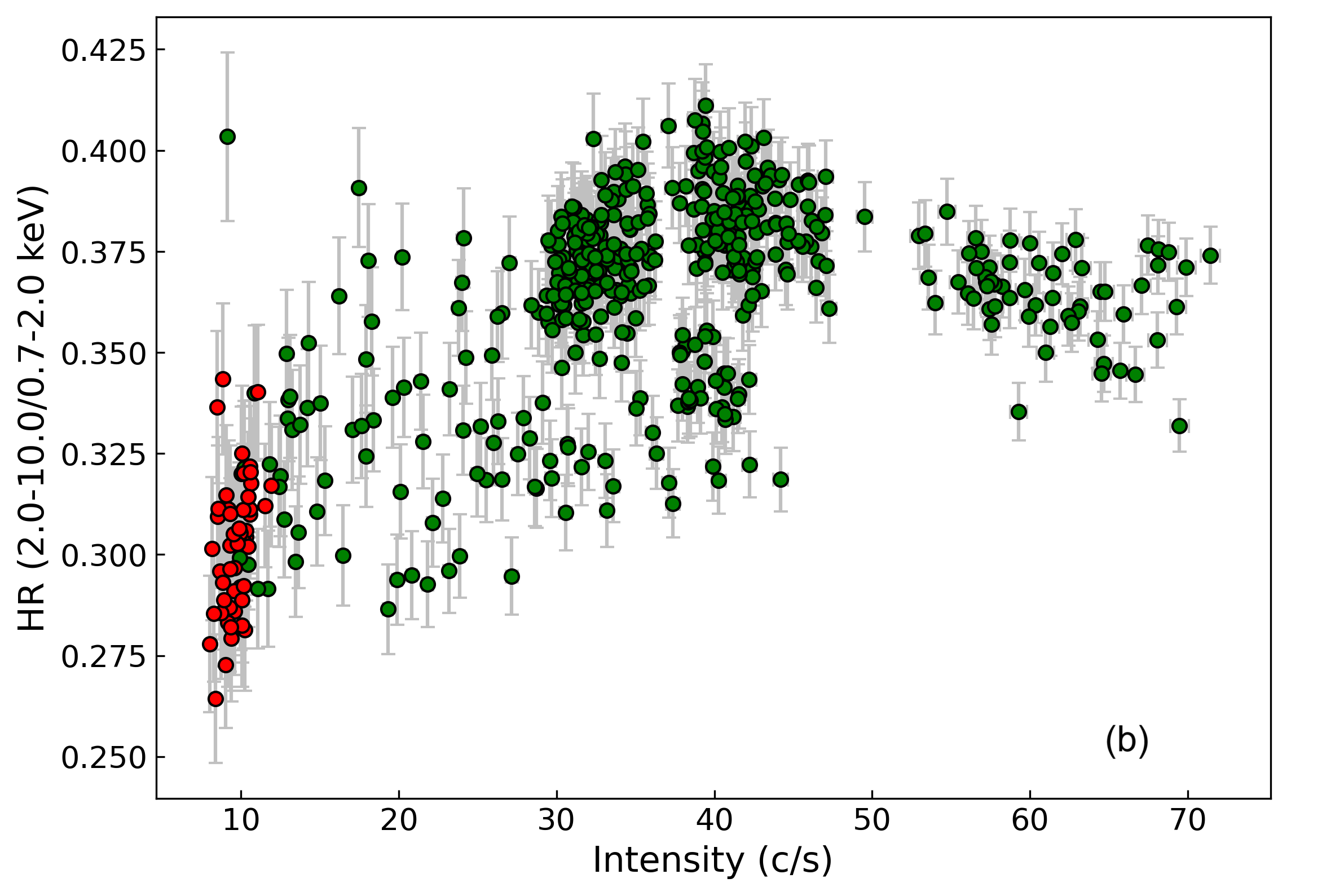}}
\caption{Panel (a): Evolution of HR of SAX~J1808 during \textit{XMM-Newton} observation of reflare phase of 2022 outburst. The HR was estimated from the (2--10~keV)/(0.7--2~keV) energy band. For comparison, the \textit{XMM-Newton} light curve is shown in blue in the background.
Panel (b): Hardness-intensity diagram (HID) of SAX~J1808 obtained using \textit{XMM-Newton} EPIC-pn observation and 200 s bins. The intensity is given in counts per second. The points corresponding to the minimum of the source X-ray flux on the light curve, between 59832.02 MJD and 59832.14 MJD, are marked in red.}
\label{fig:HR}
\end{figure}

We analyzed the X-ray spectral evolution of SAX~J1808 during the outburst's flaring stage by measuring the hardness ratio (HR), which is defined as the ratio of the count rates observed in the (2--10~keV) and (0.7--2~keV) energy bands.
Panels (a) and (b) of Fig.~\ref{fig:HR} show the variation of the HR over time and intensity, respectively.  
The source emission significantly softened during the phase of minimum X-ray count rate. The continuous decrease in the HR reached a minimum when the source flux was at its lowest. The HR then returned to its initial value, following the trend of the light curve. 

\begin{figure}
\resizebox{\hsize}{!}{\includegraphics[clip=true]{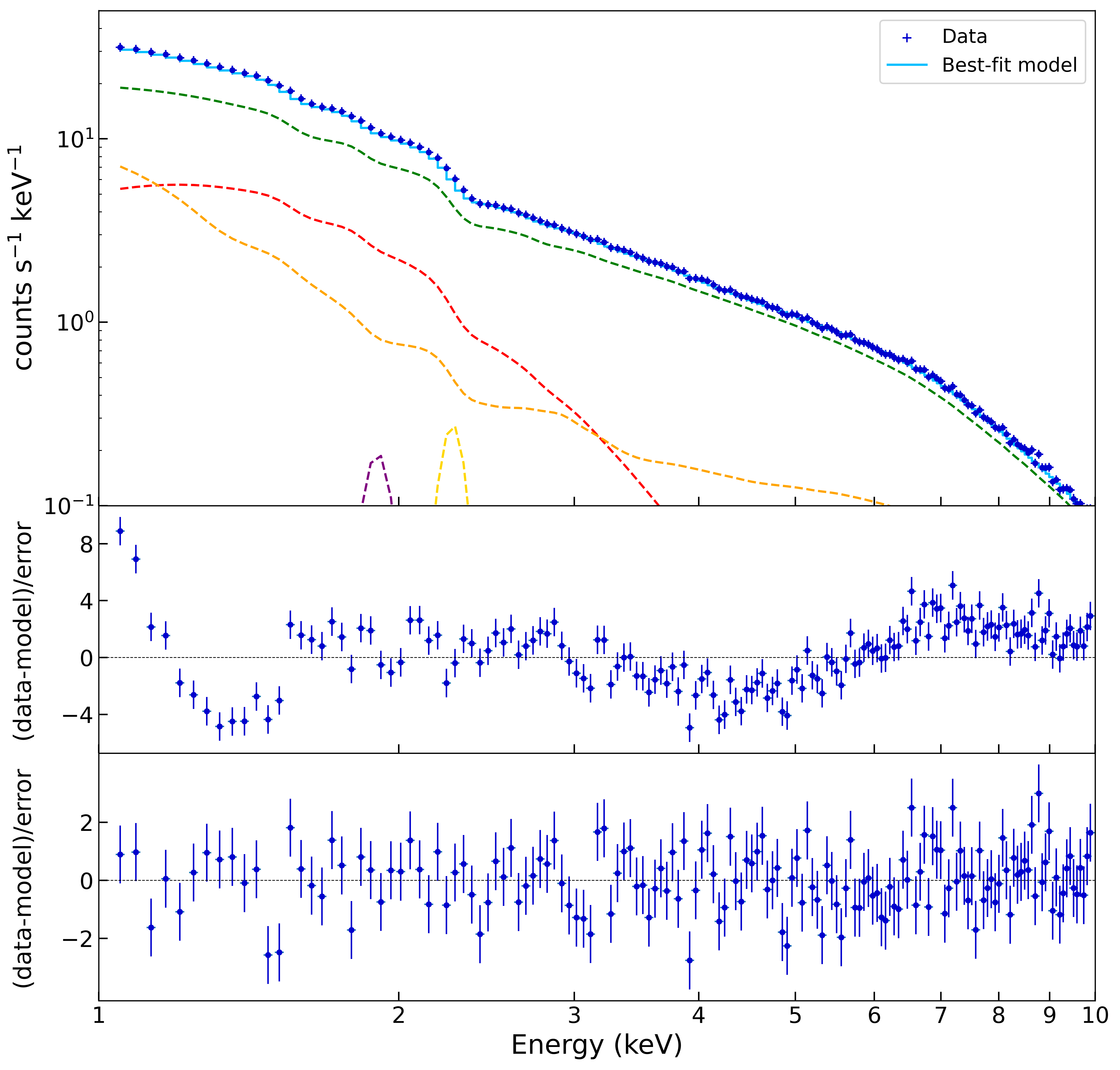}}
\caption{
\footnotesize
\textit{XMM-Newton}/EPIC-pn spectrum (1.0--10~keV) of high-flux emission from SAX~J1808 (top) and best-fit model plotted with a solid line and given in Table~\ref{tab:spectra}. The fit model is \texttt{TBabs*(bbodyrad + gaussian + gaussian + gaussian + nthComp + rdblur*rfxconv*nthComp)}. The model components are also plotted as dashed lines: the blackbody component, \texttt{bbodyrad,} in red; the Comptonization component, \texttt{nthcomp,} in green; the reflection component in orange; and the Gaussians in purple and yellow. The bottom panel displays the residuals relative to the best-fit model, while the middle panel shows the residuals from a fit performed without the reflection component.}
\label{fig:nolowspectrum}
\end{figure}

\begin{figure}
\resizebox{\hsize}{!}{\includegraphics[clip=true]{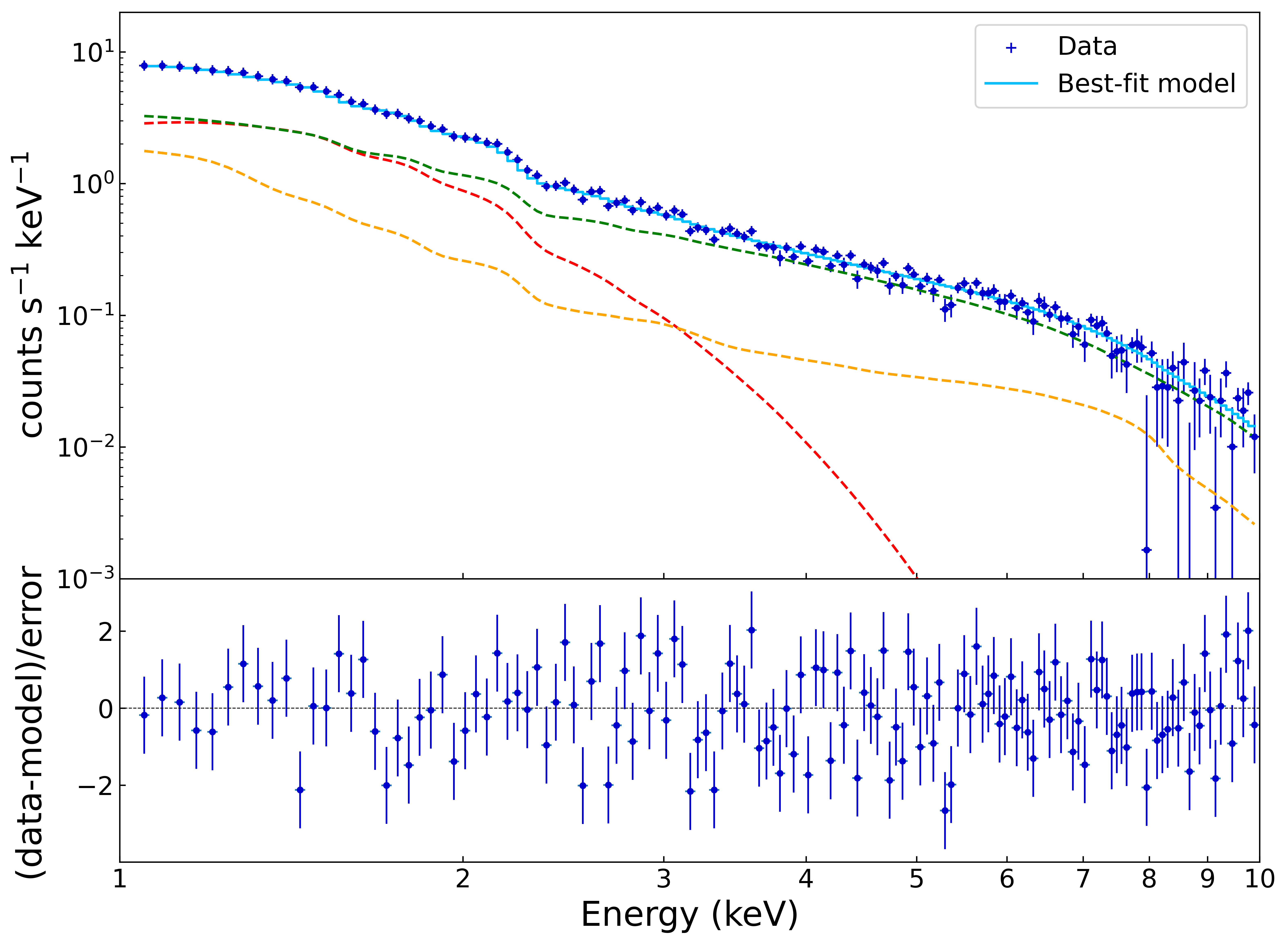}}
\caption{
\footnotesize
\textit{XMM-Newton}/EPIC-pn spectrum (1.0--10~keV) of low part of light curve of SAX~J1808 (top) and residuals with respect to best-fit model plotted with a solid line and given in Table~\ref{tab:spectra}. The fit model is \texttt{TBabs*(bbodyrad + nthComp + rdblur*rfxconv*nthComp)}. The model components are also plotted as dashed lines, with the same color-coding as in Fig.~\ref{fig:nolowspectrum}.}
\label{fig:lowspectrum}
\end{figure}

To investigate the softening of the emission, we extracted spectra considering both the low-flux phase of the light curve (between 59832.02 MJD and 59832.14 MJD; see the red band of Fig.~\ref{fig:Xlc} and corresponding red data points in Fig.~\ref{fig:HR}) and the high-flux emission, obtained by excluding the entire low-flux interval and its rise and decay phases; i.e., from MJD 59831.88 to 59832.23.
We performed the spectral analysis using the X-ray spectral fitting package \texttt{HEASARC} \texttt{XSPEC} \citep{Arnaud_1996_XSPEC} version 12.14.0. We used \texttt{TBabs} to model the interstellar photoelectric absorption, with the chemical abundances of \citet{Wilms_2000} and the cross-sections reported in \citet{Verner_1996}. To avoid the impact of calibration uncertainties at low energies, we excluded data below 1~keV from the EPIC-pn spectral analysis\footnote{\url{https://xmmweb.esac.esa.int/docs/documents/CAL-TN-0018.pdf.}}. Following previous works, we fixed the equivalent hydrogen column density at $0.21\times10^{22}\, \text{cm}^{-2}$ \citep{Papitto_2009, DiSalvo_2019}.
We first modeled the 1--10~keV spectrum observed in the high-flux interval with a blackbody component, \texttt{bbodyrad,} and a thermal Comptonization component, \texttt{nthComp} \citep{Zdziarski_1996, Zycki_1999}. This spectral decomposition is commonly used to model the broad-band continuum emission in AMSPs \citep{Poutanen_2006b, DiSalvoSanna_2022}.
The parameters of the thermal Comptonization continuum model are the asymptotic power-law photon index, $\Gamma$; the hot, Comptonizing electron temperature, $\mathrm{kT_e}$; and the seed photon temperature, $\mathrm{kT_{seed}}$. We set the $\text{inp\_type}$ parameter to zero, indicating that the seed photons are distributed as a blackbody.
Since no high-energy cut-off was observed in the \textit{XMM-Newton} spectrum, we fixed $\mathrm{kT_e}$ to 30~keV \citep[see, e.g.,][]{DiSalvo_2019}. This choice had no significant impact on the obtained results.
Using this model, we obtained an unacceptably high reduced $\chi^2$ of 6 for 146 d.o.f.. The residuals shown in the middle panel of Fig.~\ref{fig:nolowspectrum} suggest the presence of a reflection component.
We therefore convolved the \texttt{nthComp} component with the \texttt{rfxconv} model from \citet{Done_2006} and the relativistic blurring kernel \texttt{rdblur} \citep{Fabian_1989}. 
The parameters of the reflection model are the relative reflection normalization, the iron abundance relative to the solar one, the ionization parameter of the disk's $\text{log}\,\xi$, the inner- and outer-disk radii, the disk inclination, and the emissivity index, $\beta$, which describes the emissivity of the illuminated disk as a function of the emission radius ($\propto \text{r}^{\,\beta}$).
We first allowed the iron abundance to vary, but as this did not produce a significant decrease in the fit's $\chi^2$, we fixed it at 1. Given the limited spectral coverage and the weakness of the reflection features, we fixed the reflection fraction (\texttt{rel\_refl}) at $-0.62$, consistent with the value found in previous studies \citep{DiSalvo_2019}. 
We also accounted for residual emission features evident in the soft-energy part of the spectrum by introducing three \texttt{Gaussian} components. Some of these features are listed in the official \textit{XMM-Newton} user guide\footnote{\url{https://xmm-tools.cosmos.esa.int/external/xmm_user_support/documentation/sas_usg/USG/epicdataquality3.html}.} as known instrumental lines (the Si-K line at 1.84~keV and the Au-M line at 2.2~keV). We also observed an emission line at 3.5~keV, which we attributed to Ar XVIII fluorescence (see, e.g., \citealt{Pintore_2016} and \citealt{Malacaria_2025}). In some cases, the spectral resolution of the \textit{XMM-Newton}/EPIC-pn instrument was insufficient to constrain the width of the modeled lines. Therefore, we fixed the $\sigma$ parameter to 0~keV. The resulting model gave a reduced $\chi^2$ of 1.26 with 142 d.o.f.
Figure \ref{fig:nolowspectrum} shows the model and the residuals (bottom panel), while Table~\ref{tab:spectra} lists the best-fit parameters.\\
We then modeled the spectrum of the low-flux part of the light curve in the 1--10~keV energy band. We adopted the same spectral model used for the high-flux emission, but without the Gaussian components that described the low-energy emission features, which were not detectable in this spectrum, likely due to lower statistics. Figure \ref{fig:lowspectrum} shows the model and the residuals, resulting in a reduced $\chi^2$ of 1.19 with 134 d.o.f. The best-fit parameters are reported in Table~\ref{tab:spectra}.
\begin{table}[]
\renewcommand{\arraystretch}{1.1}
\centering
\caption{Best-fit parameters of \textit{XMM-Newton} EPIC-pn 1.0--10~keV spectra of SAX~J1808 during the high- and low-flux intervals. }
\label{tab:spectra}
\begin{tabular}{l c c}
\toprule
& High-flux values & Low-flux values\\
\midrule
\scshape{TBabs} & & \\
$N_{\rm{H}}$ [$10^{22}$ cm$^{-2}$] & $0.21^{(a)}$ & $0.21^{(a)}$ \\
\midrule
\scshape{bbodyrad} & & \\
 $kT_{\rm{BB}}$ [keV] & $0.399^{+0.003}_{-0.004}$ & $0.362^{+0.011}_{-0.013}$ \\
 $R_{\rm{BB}}$ ($d_{3.5}$ km) & $3.26^{+0.10}_{-0.07}$ & $2.86\pm0.14$ \\
 $F_{0.5-10}$$^{b}$ & $0.218^{+0.014}_{-0.018}$ & $0.102^{+0.010}_{-0.013}$ \\
\midrule
\scshape{Gaussian lines} & & \\
Si K-edge $E_{\rm{line}}$ [keV] & $1.91\pm0.02$ & -- \\
 $\sigma$ [keV] & $0^{(a)}$ & -- \\
 Norm [$10^{-5}$] & $3.3^{+0.7}_{-0.9}$ & -- \\
 Au M-edge $E_{\rm{line}}$ [keV] & $2.265^{+0.009}_{-0.007}$ & -- \\
 $\sigma$ [keV] & $0^{(a)}$ & -- \\
 Norm [$10^{-5}$] & $6.5\pm0.7$ & -- \\
 Ar XVIII $E_{\rm{line}}$ [keV] & $3.49\pm0.05$ & -- \\
 $\sigma$ [keV] & $0.37\pm0.07$ & -- \\
 Norm [$10^{-5}$] & $9\pm2$ & -- \\
\midrule
\scshape{nthComp} & & \\
 $\Gamma$ & $2.04^{+0.05}_{-0.04}$ & $2.06^{+0.07}_{-0.04}$ \\
 $kT_{\rm{e}}$ [keV] & $30^{(a)}$ & $30^{(a)}$ \\
 $kT_{\rm{seed}}$ [keV] & $<0.11$ & $<0.15$ \\
 Redshift & $0^{(a)}$ & $0^{(a)}$ \\
 Norm & $0.0300\pm0.001$ & $0.0051^{+0.0004}_{-0.0003}$ \\
 $F_{0.5-10}$$^{b}$ & $1.39^{+0.03}_{-0.04}$ & $0.23\pm0.04$ \\
\midrule
\scshape{Reflection} & & \\
 $\beta$ & $-3.02^{+0.10}_{-0.20}$ & $-3.08^{+0.4}_{-1.5}$ \\
 $R_{\rm{in}}$ [$R_{\rm{g}}$]$^{c}$ & $<6.7$ & $<14.7$ \\
 $R_{\rm{out}}$ [$R_{\rm{g}}$]$^{c}$ & $1000^{(a)}$ & $1000^{(a)}$ \\
 Incl. [deg] & $58^{+2}_{-1}$ & $59^{+5}_{-6}$ \\
 Refl. frac. & $-0.62^{(a)}$ & $-0.62^{(a)}$ \\
 Fe abund. [solar] & $1^{(a)}$ & $1^{(a)}$ \\
 $\log{\xi}$ & $2.32\pm0.01$ & $2.7\pm0.2$ \\
 $F_{0.5-10}$$^{b}$ & $0.372^{+0.012}_{-0.014}$ & $0.0901^{+0.005}_{-0.007}$ \\
\midrule
\scshape{Total} & & \\
$F_{0.5-10}$$^{b}$ & $1.989^{+0.006}_{-0.04}$ & $0.424^{+0.014}_{-0.010}$ \\
\midrule
 $\chi^2$/d.o.f & 179.3/142 & 159.4/134 \\
\bottomrule
\end{tabular}
\tablefoot{The fit models are \texttt{TBabs*(bbodyrad + gaussian + gaussian + gaussian + nthComp + rdblur*rfxconv*nthComp)} for the high-flux state and \texttt{TBabs*(bbodyrad + nthComp + rdblur*rfxconv*nthComp)} for the low-flux state. Uncertainties are given at a 1$\sigma$ confidence level. \\
\tablefoottext{$a$}{Parameter kept frozen during the fit.}
\tablefoottext{$b$}{Unabsorbed flux in the 0.5--10~keV band and reported in units of $10^{-10}$ erg cm$^{-2}$ s$^{-1}$.}
\tablefoottext{$c$}{$R_{\text{g}} = GM/c^2$ is the gravitational radius.}
}
\end{table}

\section{Spectral energy distribution} \label{sec:sed}
\begin{figure}
\resizebox{\hsize}{!}{\includegraphics[clip=true]{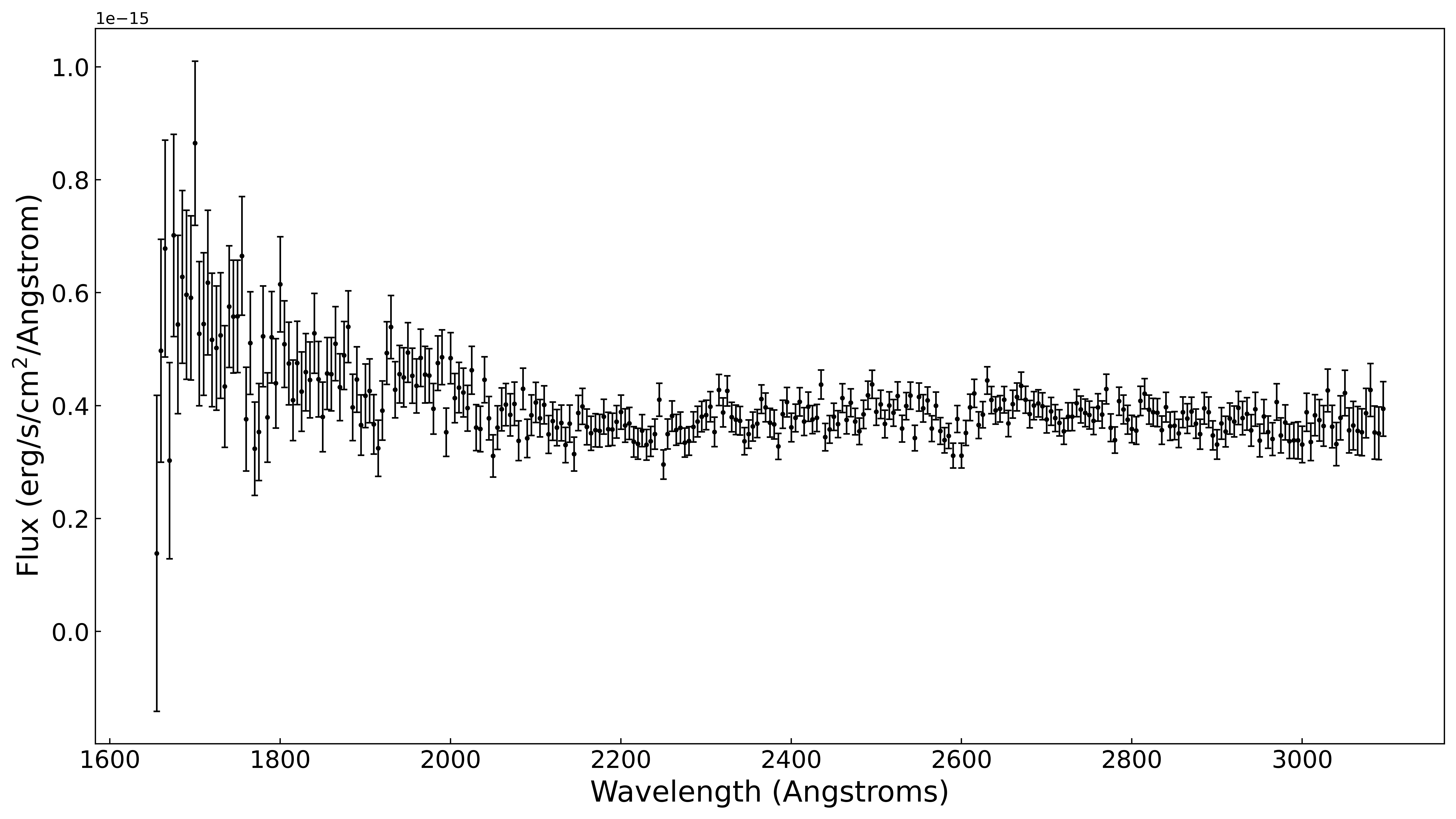}}
\caption{
\footnotesize
UV spectrum of \textit{HST}/STIS observation of SAX~J1808 performed on September 10, 2022 and binned at $5\,\AA$ in the $1650-3100\, \AA$ wavelength range.}
\label{fig:spettrouv}
\end{figure}

Figure~\ref{fig:spettrouv} shows the UV spectrum of SAX~J1808, covering the wavelength range of $1650-3100\, \AA$. 
The only notable feature is a possible weak absorption line near $2600\, \AA\,$, most likely due to Fe II.
We corrected the UV spectrum for interstellar extinction using the empirical relation $A_V = N_{\rm{H}}/(2.87 \pm 0.12) \times 10^{21} \, \text{cm}^{-2}$ \citep{Foight_2016}, where $N_{\rm{H}} = 2.1 \times 10^{21} \, \text{cm}^{-2}$ represents the hydrogen column density along the line of sight to SAX~J1808 \citep{DiSalvo_2019}, which gives $A_V \simeq 0.73$\,mag. We obtained a corresponding reddening of $E(B-V)\simeq0.27$ using the extinction coefficients of \citet{Schlafly_2011} for $R_V=3.1$. We then applied the extinction law of \citet{Fitzpatrick_1999} to compute $A_\lambda$ across the STIS spectral range and corrected each wavelength bin with the corresponding extinction value to derive the dereddened spectrum.
We numerically integrated the spectrum with the trapezoidal rule, which approximates the area under the curve by dividing it into small trapezoids, obtaining a UV-flux value of $F_{\rm{UV}}=(3.7 \pm 0.7)\times 10^{-12}\,\mathrm{erg\, cm^{-2}\, s^{-1}}$.
Considering the measured UV pulse amplitude (see Sect.~\ref{sec:UV}), the coherent UV pulses detected during the \textit{HST}/STIS observation showed a pulsed luminosity of $L_{\rm{pulsed}}^{\rm UV} = 0.019 \, L_{\rm{UV}} \approx (1.1\pm0.3) \times 10^{32} \,\text{erg s}^{-1}$. 

\begin{figure}
\resizebox{\hsize}{!}{\includegraphics[clip=true]{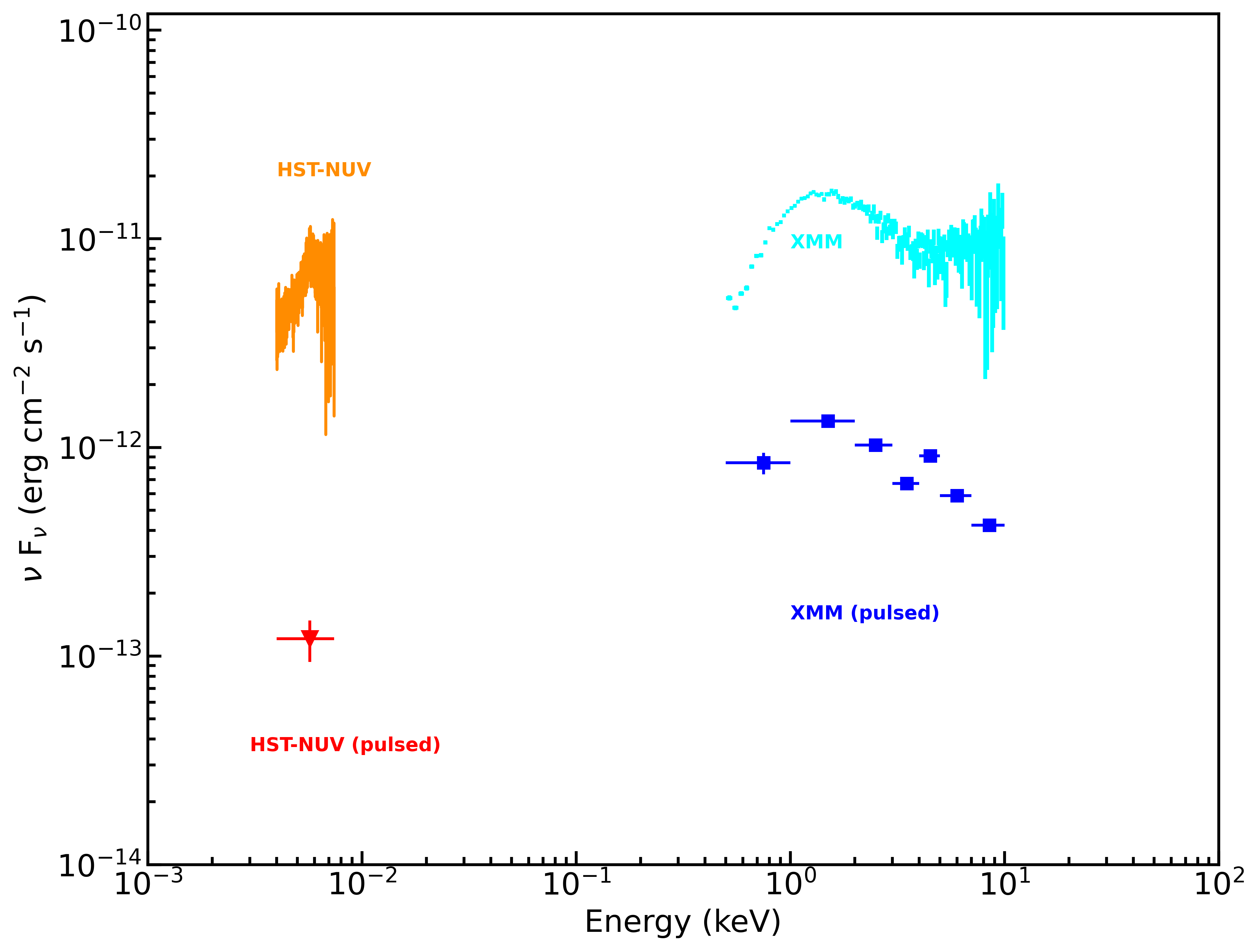}}
\caption{
\footnotesize
Unabsorbed SED of SAX~J1808 from UV to X-rays obtained using simultaneous \textit{XMM-Newton}/\textit{HST }(STIS) observations performed during the 2022 outburst reflare phase (September 10, 2022). The \textit{HST} spectrum is shown in orange from 165 to 310 nm. The pulsed UV flux is plotted with a red triangle. The total 0.5--10~keV X-ray fluxes observed by \textit{XMM-Newton} are plotted using light blue points. The pulsed X-ray fluxes, plotted as dark blue squares, are calculated over the 0.5--1, 1--2, 2--3, 3--4, 4--5, 5--7, and 7--10 keV energy bands.}
\label{fig:sed}
\end{figure}

Figure~\ref{fig:sed} shows the spectral energy distribution (SED) of the total and pulsed emission of SAX~J1808 in the UV and X-ray bands during the September 10, 2022 simultaneous observations.
To evaluate the X-ray pulsed flux simultaneously with the \textit{HST} observations, we measured the r.m.s. amplitude in seven energy bands (Fig.~\ref{fig:ampvsenergy}) and multiplied these values for the unabsorbed X-ray fluxes calculated over the same energy ranges. We then scaled the resulting values by the ratio between the mid-point energy of the interval and the width of the energy interval to convert the pulsed X-ray fluxes into $\nu F_{\nu}$ units.

\section{Discussion}
\subsection{Phase shifts and hot-spot displacements}\label{sec:timingandshifts}
We analyzed an uninterrupted \textit{XMM-Newton} observation of the AMSP SAX~J1808 performed during the final reflaring stage of the 2022 outburst. 
X-ray coherent pulsations were detected throughout the observation, providing strong evidence for channeled accretion onto the magnetic poles of the NS at very low X-ray luminosity. 
The 0.5--10~keV luminosity measured during the phase of minimum observed X-ray flux (red band in Fig.~\ref{fig:Xlc}, Table~\ref{tab:spectra}) was $L_{\text{X(low)}\,0.5-10} \simeq 6.21^{+0.20}_{-0.15} \times 10^{34} \, d\mathrm{^2_{3.5}\,erg \, s^{-1}}$.
This value is on the order of the lowest X-ray luminosity at which pulsations of SAX~J1808 were previously observed (around a few $\times 10^{34}\,\mathrm{erg \,s^ {-1}}$; \citealt{Hartman_2008, Hartman_2009, patruno_2009b, Bult_2019, Archibald_2015}). 

To estimate the location of the inner-disk radius during the observation, we used the relation $R_{\rm{m}} \simeq 0.5\, \mu^{4/7}\dot{M}^{-2/7}(2GM)^{-1/7}$ \citep{Pringle_1972, Lamb_1973, Campana_2018}, where $\mu=\mathrm{BR_*^3}\sim 10^{26}\, \mathrm{G\,cm^{3}}$ is the magnetic moment measured from the secular spin-down rate \citep{Hartman_2008}, and $\dot{M}$ is related to the luminosity as $\dot{M}=L_X R_*/GM_*$.
We estimated the bolometric 0.1--100.0~keV X-ray luminosity during the low- and high-flux parts of the observations by extrapolating the best-fitting spectral model from the 1.0--10~keV range. We obtained $L_{\text{X(low)}\,0.1-100}~\simeq 0.91^{+0.19}_{-0.06} \times~10^{35} \, d^2_{3.5}\,\mathrm{erg \, s^{-1}}$ and $L_{\text{X(high)}\,0.1-100} \simeq 5.77^{+0.20}_{-0.12} \times \mathrm{10^{35} \, erg \, s^{-1}}$, respectively. The corresponding estimates of the inner-disk radius are $R_{\rm{m(low)}}\simeq~36.55^{+0.40}_{-0.13}\, \text{km}$ and $R_{\rm{m(high)}}\simeq~21.6^{+0.4}_{-0.2}\, \text {km}$. Note that these uncertainties are likely underestimated, as they do not include contributions from uncertainties of the magnetic-field strength, mass, and radius.
For comparison, the corotation radius of SAX~J1808 is $R_{\rm{co}}=31\,m\mathrm{^{1/3}_{1.4}\,km}$, where $m_{\rm{1.4}}$ is the NS mass in units of $1.4\,M_{\odot}$.
The observation of X-ray pulsations in the low-flux interval confirms that channeled accretion continues even when the mass-accretion rate is low and that the inner disk remains close to the corotation radius, likely in a weak propeller regime \citep{patruno_2016}. Notably, the X-ray pulsations showed a larger amplitude at the lowest observed fluxes (second panel of Fig.~\ref{fig:residuals}). This behavior has been observed in several AMSPs, where the pulsed fraction tends to increase as the accretion rate drops (see, e.g., IGR~J17379-3747, MAXI~J1957+032, MAXI~J1816-195, IGR~J17498-2921; \citealt{Bult_2019b}, \citealt{Sanna_2022c}, \citealt{Bult_2022}, and \citealt{Illiano_2024}).

A phase-coherent timing analysis revealed a clear phase jump of $\Delta \phi \simeq 0.4$ cycles, coincident with a decrease by a factor of three of the X-ray luminosity to the lowest value measured during our observation ($L_{\text{X(low)}}\simeq10^{35}\,\mathrm{erg\,s^{-1}}$). Timing noise is known to affect the phases observed from SAX~J1808 \citep{Hartman_2008, Hartman_2009} and other AMSPs \citep{Patruno_2009}.
\citet{Burderi_2006} observed a phase shift of the opposite sign and roughly half an order of magnitude ($\Delta\phi\simeq -0.2$) during the 2002 outburst; this was simultaneous to a rapid drop of the X-ray flux and before the beginning of the reflaring phase. Several models have been proposed to explain such changes.
\citet{Poutanen_2009} and \citet{Ibragimov_2009} interpreted the phase shift and the appearance of a secondary peak at the end of the 2002 outburst in terms of the opening of the view to the antipodal spot due to the recession of the inner disk when the mass-accretion rate decreased.
A related interpretation involves changes in the disk-magnetosphere coupling and emission geometry \citep{Kajava_2011}.
However, in our observation, no secondary peak appeared when the X-ray flux dropped. Instead, the amplitude of the second harmonic became weaker, while the first harmonic became twice as strong compared to the higher flux level (see Figs. \ref{fig:Xpulseprofiles} and \ref{fig:residuals}).
Assuming the drop of the X-ray flux causes an increase in $R_{\rm{in}}$, the lack of a strong secondary peak suggests that the disk was already sufficiently far from the NS to avoid significantly occulting the secondary pole. Indeed, the observed X-ray flux ($\lesssim 1.5\times10^{-10}$ erg cm$^{-2}$ s$^{-1}$; 0.5--10 keV) was lower than the typical flux at which the transition to the rapid drop occurred in 2002 ($\gtrsim 10^{-9}$ erg cm$^{-2}$ s$^{-1}$).
An alternative explanation involves scattering of the hot-spot emission in the accretion funnel \citep{Ahlberg_2024}, but this effect is only relevant at higher accretion rates ($\dot{M} \gtrsim 10^{-10}\, \mathrm{M_\odot\,yr^{-1}}$) than those observed in our observation.

Another possible explanation for the observed phase shifts in AMSPs involves spot movement in the azimuthal direction \citep{Lamb_2009, Kulkarni_2013}. In particular, variations in the hot-spot longitude directly affect the pulse arrival times. Such drifts can result from changes in the mass-accretion rate. As the accretion disk recedes, the inflowing matter couples to different magnetic-field lines and is channeled to slightly displaced regions on the NS surface.
The largest phase variation we observed in the \textit{XMM-Newton} data taken during the reflaring stage of the 2022 outburst corresponds to a shift in the hot-spot longitude of $\simeq 100^{\circ}$. \citet{Patruno_2009} first proposed the existence of a correlation between the observed pulse phase and the X-ray flux, based on the improvement it produced in the description of the phase evolution of a sample of AMSPs. 
Thus, to account for the phase shifts observed in our analysis, we added a phase-flux correlation term to the constant frequency model, $R_{\rm{flux}}(t)=b\,(F_{\rm{X}}/F_0)^\Lambda$, as in \citet{Bult_2020}. This addition yielded a significant improvement in the description of the temporal evolution of the observed pulse phases.
The value of $\Lambda=-0.17(9)$ we obtained agrees with the value suggested by numerical simulations of accretion onto a rapidly rotating NS \citep{Kulkarni_2013}. These simulations suggest that the azimuthal position of the hot spot relative to the magnetic pole scales proportionally with the magnetospheric radius, $\delta \phi \propto R_{\text{m}}$. Assuming $R_{\text{m}}\propto \dot{M}^{\Lambda}$ and $\dot{M}$ to be proportional to the bolometric flux, we obtain $\delta \phi \propto R_{\text{m}} \propto F_X^{\Lambda}$. The value we found is compatible with the index $\Lambda=-1/5$, which we found by defining the inner-disk radius as the point where the inflowing matter is forced into corotation with the NS magnetosphere \citep{Spruit_1993, Dangelo_2010}. An even steeper dependence of approximately the usual definition of the Alfvén radius, $\Lambda=-2/7$, cannot be excluded, however.

The contour plot in the $(b,\Lambda)$ plane reveals a correlation between these two parameters, $b\cdot\Lambda\simeq \text{const}$ (dashed black line in Fig.~\ref{fig:chi2curve}). Such a correlation likely arose because the fit was primarily driven by the step-like variation of $\Delta \phi \simeq 0.4$ in response to a variation in the flux $\Delta F_{\rm{X}} \simeq 50\,\text{c\,s}^{-1}$, with a similar shape.
Consequently, the fit was mainly sensitive to the ratio $\Delta \phi / \Delta F_{\rm{X}} \simeq 8\times10^{-3}\, \text{s}$, which approximates the derivative of the functional form of the phase-flux correlation, $d\phi/dF_{\rm{X}}=b\Lambda\,\cdot F_0\cdot(F_{\rm{X}}/F_0)^{\Lambda-1}\sim b\cdot \Lambda$. Observations of a few phase jumps of variable amplitude, simultaneous to flux variations of different magnitudes, would provide a stronger test of the model and likely remove the degeneracy. 

Analysis of \textit{NICER} observations from the 2019 and 2022 outbursts yielded consistent values of $\Lambda$ within the uncertainties.
Whereas for the 2019 dataset \citet{Bult_2020} fixed the $\Lambda$ index to the theoretically expected value of $-1/5$, we treated $\Lambda$ as a free parameter, obtaining $\Lambda = -0.14(5)$.
Our analysis of 2022 data gave a value of $\Lambda = -0.37(14)$, which differs from that reported by \citet{Illiano_2023} ($\Lambda = -0.81(12)$). Such a difference is due to our decision not to consider the phase evolution after the first long gaps in data, as these would have prevented us from reliably approximating the flux with a spline function.

When the accretion disk recedes toward the end of an outburst, the spot latitude is also expected to change as matter is channeled toward the NS surface by magnetic lines that land closer to the magnetic pole. Such variations modify the amplitude of the harmonic components of the pulsed profile \citep{Poutanen_2006, Lamb_2009}.
In particular, if the hot spot is close to the stellar spin axis, modest changes in its latitude can produce significant variations. The effect of such changes on the observed pulsed fraction depends on the initial spot latitude (see Fig.~1 from \citealt{Lamb_2009} and Fig.~6 from \citealt{Poutanen_2006}). If only a single spot is visible, a decrease in the spot latitude generally reduces the pulsed fraction. On the other hand, if two antipodal spots are visible and the primary spot latitude is $\gtrsim 45^\circ$, its decrease might produce an increase in the fractional amplitude.
Simultaneous to the flux decrease and the phase shift discussed above, we observed an increase in the amplitude of the fundamental frequency up to $\simeq$10\% (second panel of Fig.~\ref{fig:residuals}).
Such an increase in the pulsed fraction might thus be interpreted in terms of a change in the spot latitudes caused by a receding inner disk, provided that SAX~J1808 is an approximately orthogonal rotator whose pulse profile arises from two spots.
In such a geometry, the presence of two nearly antipodal hot spots is also expected to enhance the polarimetric signature. This effect was recently observed by IXPE in the orthogonal rotator GRO~J1008$-$57 \citep{Tsygankov_2023}, suggesting that similar measurements of SAX~J1808 could provide additional constraints on its geometry.

\subsection{UV pulsations}
We presented a timing analysis of an \textit{HST}/STIS UV observation performed simultaneously with the interval of minimum X-ray flux observed by \textit{XMM-Newton} (Fig.~\ref{fig:Xlc}). We confirmed the significant detection of UV pulsations, with a pulsed luminosity of $L_{\text{pulsed}}^{\text{UV}} = 1.1(3) \times 10^{32} \, \text{erg} \, \text{s}^{-1}$ consistent with that observed during the previous accretion event of SAX~J1808 in 2019 \citep{Ambrosino_2021}. 
\begin{table}
\centering
\caption{Total and pulsed luminosities in X-ray and UV bands computed for the 2019 and 2022 outbursts.}
\begin{tabular}{lcc}
\toprule
& 2019 & 2022 \\
\midrule
$L_{\rm X(tot)}$ $[\mathrm{erg\,s^{-1}}]$ & $1.19(1)\times 10^{35}$ & $6.21_{-0.15}^{+0.20}\times 10^{34}$ \\ 
$L_{\rm X(pulsed)}$ $[\mathrm{erg\,s^{-1}}]$ & $7(1)\times10^{33}$ & $4.4(4)\times 10^{33}$ \\
$L_{\rm UV(tot)}$ $[\mathrm{erg\,s^{-1}}]$ & $5(1)\times10^{33}$ & $5(1)\times10^{33}$ \\
$L_{\rm UV(pulsed)}$ $[\mathrm{erg\,s^{-1}}]$ & $1.5(5)\times10^{32}$  & $1.1(3) \times 10^{32}$ \\
\bottomrule
\end{tabular}
\label{tab:luminosities}
\end{table} 
In both the 2019 and 2022 outbursts, UV pulsations were detected during the reflaring phase (see our Fig.~\ref{fig:Xlc} and Fig. 1 of \citealt{Ambrosino_2021}), across an X-ray luminosity range spanning a factor of $\sim 2$ (see Table \ref{tab:luminosities}). 
The values of the pulsed X-ray and UV luminosity we observed simultaneously in the 2022 outburst (see Table~\ref{tab:luminosities}) give a ratio of $L_{\rm pulsed}^{\rm X}/L_{\rm pulsed}^{\rm UV}=41\pm13$. Applying the same analysis to the non-simultaneous 2019 observations yields a similar ratio of $48 \pm 18$.

Standard mechanisms hardly explain the very bright, pulsed optical and UV flux of SAX~J1808. In fact, the observed values imply a brightness temperature of $>1$\,MeV for a 10\,km-wide NS at the distance of SAX~J1808, which can be safely excluded given the flux observed in the X-ray band \citep{Ambrosino_2021}. On the other hand, the thermal emission observed at soft X-rays is expected to produce a 165--310~nm UV luminosity of
\begin{equation}
    L_{\text{thermal}}=A \int_{\nu_1}^{\nu_2}(2\pi k_{\text{B}} T_{\rm{bb}}\nu^2/c^2)d\nu, 
\end{equation}
where  $k_{\rm{B}}$ is the Boltzmann constant, $kT_{\rm{bb}}$ is the blackbody temperature, $A\approx \pi R_*^2(R_*/R_{\rm{co}})$ is the hot-spot area of the accreting polar cap ($\sim 100\, \mathrm{km^2}$ if the whole NS is involved; \citealt{Frank_2002}), and $\nu_1$ and $\nu_2$ represent the lower and upper frequency limits of the UV band. Using $kT_{\rm{bb}}=0.362\,\text{keV}$ (Table~\ref{tab:spectra}), the expected UV luminosity is $\sim 8\times 10^{27}\, \mathrm{erg\,s^{-1}}$, which is four orders of magnitude lower than observed.
On the other hand, cyclotron emission from hot electrons in the post-shock region of the accretion columns of a $3.5\times10^8$~G magnetic-field pulsar can account for a UV luminosity up to $\simeq 6\times 10^{29}\,\mathrm{erg\,s^{-1}}$ \citep{Ambrosino_2021}, which is lower by a factor of $150$ than the value reported here. Even assuming extreme beaming, it seems highly unlikely that such processes can account for the brightness of the UV pulsations observed from SAX~J1808.

The transitional millisecond pulsar PSR~J1023+0038 is the only other millisecond pulsar known to show both UV and optical pulsations \citep{Ambrosino_2017, Papitto_2019, Ambrosino_2021, Jaodand_2021, Miraval_2022}. These signals were detected when the source was in an X-ray-faint ($L_X\simeq \mathrm{a\,few} \times 10^{33}\,\mathrm{erg\, s^{-1}}$) accretion-disk regime. The SED of the pulsed flux of PSR~J1023+0038 was compatible with a power-law relation connecting the optical and UV to the X-ray band \citep{Papitto_2019, Miraval_2022}. Recently, \citet{Baglio_2024} showed that the SED of polarized emission also follows a compatible trend. Synchrotron emission from electrons accelerated where the rotation-powered electromagnetic wind of the pulsar interacts with the mass inflow has been proposed to explain the high pulsed luminosity observed at optical and UV energies (\citealt{Papitto_2019, Veledina_2019}; see also \citealt{Illiano_2023b, Baglio2023}). 
Unlike the case of PSR~J1023+0038, the simultaneous estimates of the pulsed X-ray and UV fluxes of SAX~J1808 cannot be connected by a single power law (see Fig.~\ref{fig:sed}), arguing against an interpretation in terms of a standard synchrotron-emission process. Most importantly, the observed X-ray luminosity of SAX~J1808 ($L_X\simeq \mathrm{a\,few} \times 10^{35}\,\mathrm{erg\, s^{-1}}$) ensures that the system is accretion powered and that high-density plasma engulfs the magnetosphere. 
\citet{Ambrosino_2021} suggested the possibility of a coexistence or a fast alternation between optical and UV pulsed radiation produced by electrons accelerated by a rotation-powered pulsar and X-ray pulsed emission driven by mass accretion onto the polar caps of the NS. While future theoretical investigation will assess the feasibility of such an unusual scenario, an immediate explanation of the high pulsed luminosity of SAX~J1808 that does not rely on strong beaming of the pulsed flux is lacking.

\subsection{Evolution in the X-ray spectrum}
A study of the evolution of the HR of the source during the \textit{XMM-Newton} observation revealed a softening of the emission corresponding to the minimum of the source's X-ray flux. To investigate the origin of this behavior, we performed spectral analysis of both the high- and low-flux intervals of the light curve. The spectra were fit with a model consisting of a blackbody component, a Comptonization component, and a reflection component. 
Most of the best-fit parameters obtained for the high- and low-flux intervals are consistent at the $3\sigma$ confidence level. The normalization of the Comptonization component was the only parameter that decreased significantly in the low-flux interval. The normalization of the \texttt{bbodyrad} component, instead, shows only a slight variation, barely reaching the $3\sigma$ confidence level.
The ratio between the flux in the Comptonization and blackbody components (see Table~\ref{tab:spectra}) changed from $6.4^{+0.5}_{-0.4}$ in the high-flux state to $2.3\pm0.4$ in the low-flux state. 
Such a behavior could be related to the different physical origins of the two components: the thermal blackbody emission likely originates from a fraction of the NS surface; while the Comptonization emission arises from up-scattering of these same seed photons by electrons in the accretion columns \citep{Poutanen_2006b}.
A decrease in mass accretion rate might then reduce the efficiency of the Comptonization process faster than it affects the thermal blackbody emission, possibly leading to a softer spectrum at lower fluxes.

The parameters measured for the soft and reflection components can provide constraints on the geometry of the emitting regions. The normalization of the \texttt{bbodyrad} component can be used to estimate the radius of the emitting blackbody region, $R_{\rm bb}= \sqrt{Norm_{\rm{bbodyrad}}} \cdot D_{10}$, where $D_{10}=0.35$ is the distance to the source in units of 10 kpc \citep{galloway_2006}. We thus obtained $R_{\rm{bb}}=3.26^{+0.10}_{-0.07}\, \text{km}$ for the high-flux emission spectrum and $R_{\rm{bb}}=2.86 \pm 0.14\,\text{km}$ for the low-flux part of the light curve.
The blackbody component can be interpreted as unscattered emission from a portion of the NS surface. Interestingly, the region appears smaller at lower flux values, possibly suggesting a shrinkage of the emission region.
The best-fit parameters of the \texttt{rdblur} convolution model of the high-flux spectrum returned an estimate for the inner-disk radius of $R_{\rm in}<6.7\, R_{\rm g}$ (in units of the gravitational Schwarzschild radius $R_{\rm g} = GM/c^2$), and $i=(58^{+2}_{-1})^{\circ}$ for the inclination. The inner-disk radius is equivalent to values of $<14\,\text{km}$ for a NS mass of $\mathrm{1.4\, M_{\odot}}$, which is consistent with the inner disk being truncated before the NS surface and inside the corotation boundary. However, the $R_{\rm in}$ value inferred from the reflection model appears smaller than expected if the disk is truncated near the corotation radius, as might be suggested by the observed low flux (see Sect. \ref{sec:timingandshifts} for further discussion).
Both the inner radius and inclination are consistent with previous estimates based on the relativistic broadening of the reflection features of this source \citep{Papitto_2009, Cackett_2009, DiSalvo_2019}.
An optical estimate based on quiescent light-curve modeling yields a lower inclination of $50^{+6}_{-5}$ degrees \citep{Wang_2013}. Such a low inclination would imply an unusually low NS mass to satisfy the companion’s radial-velocity constraints. This would, in turn, exacerbate the tension with the higher inclination required by the X-ray reflection analysis (see Sect. 4.3 of \citealt{DiSalvo_2019} for further discussion).
The uncertainties on the reflection parameters are coarser. This hampers the detection of a change in the inner-disk radius, as suggested by the variation in pulse amplitude and phase. 

\section{Conclusions}
We investigated the temporal and spectral properties of simultaneous X-ray and UV pulsations from the AMSP SAX~J1808. We analyzed \textit{XMM-Newton} and \textit{HST} observations obtained at low accretion rates during the final reflaring stage of its 2022 outburst. The main results are as follows.
\begin{enumerate}
\item We detected significant X-ray pulsations down to a luminosity of $L_{\rm{X}\,0.5-10} \simeq 6.21^{+0.20}_{-0.15} \times 10^{34} \, d^2_{3.5}\,\mathrm{erg \, s^{-1}}$ in the 0.5--10~keV band. This value is of the order of the lowest X-ray luminosities at which X-ray pulsations have ever been detected in this source (approximately a few $10^{34}\,\mathrm{erg\,s^{-1}}$; \citealt{Bult_2019}).
\item \textit{XMM-Newton} data revealed significant variations in both pulse phase and amplitude. In particular, a distinct phase jump of approximately 0.4 was observed, coinciding with an amplitude increase up to $\sim$10\% and a decrease in X-ray flux to the lowest luminosity reached during our observation. We interpret these shifts as the drift of emission regions in longitude and latitude on the NS surface as a function of the mass-accretion rate. 
\item To take into account the phase shifts, we modeled the phase delays using the flux-adjusted model \citep{Bult_2020}, obtaining an index of correlation between the phases and the X-ray flux of $\Lambda=-0.17(9)$. This value matches the expected scaling of the magnetospheric radius with the accretion rate, which is predicted to be $-1/5$ from numerical simulations of accretion onto a rotating NS \citep{Kulkarni_2013}.
\item During the X-ray flux minimum, simultaneous \textit{HST} observations confirmed significant UV pulsations. The pulsed UV luminosity ($L_{\text{pulsed}}^{\text{UV}} = 1.1(3) \times 10^{32} \, \text{erg} \, \text{s}^{-1}$) was consistent with that observed during the 2019 outburst. In addition, the pulsed X-ray-to-UV luminosity ratio remained consistent across the two outburst events. The SED of the pulsed emission cannot be explained by a single power-law or thermal model, challenging standard emission scenarios. 
\item Spectral analysis of high- and low-flux intervals of the \textit{XMM-Newton} observation revealed a softening of the emission during the minimum of the X-ray source flux, associated with a decrease in the Comptonization component. The measured inclination ($i=(58_{-1}^{+2})^{\circ}$) and the inner-disk radius ($<14\,\text{km}$, for a NS mass of $\mathrm{1.4\, M_{\odot}}$) remain consistent with previous measurements, supporting a truncated disk within the corotation radius.
\end{enumerate}

\begin{acknowledgements}
We thank the referee for useful comments.
This work is based on observations obtained with \textit{XMM-Newton}, an ESA science mission with instruments and contributions directly funded by ESA Member States and NASA; the NASA/ESA Hubble Space Telescope obtained from the Space Telescope Science Institute (program GO/DD-17245), which is operated by the Association of Universities for Research in Astronomy, Inc., under NASA contract NAS 5–26555; NASA through the \textit{NICER} mission and the Astrophysics Explorers Program.
This work was supported by INAF (Research Grant ‘Uncovering the optical beat of the fastest magnetized neutron stars (FANS)’ and ‘Polarized X-rays from an accreting millisecond pulsar: a pathway to the equation of state of neutron stars (PULSE-X)’, PI: Papitto), the Italian Ministry of University and Research (MUR PRIN 2020) Grant 2020BRP57Z, ‘Gravitational and Electromagnetic-wave Sources in the Universe with current and next-generation detectors (GEMS)’, PI: Astone) and the Fondazione Cariplo/Cassa Depositi e Prestiti (Grant 2023-2560 ‘Taking the optical pulse of the quickest spinning Neutron Stars: a pilot exploratory study (SPES)’, PI: Papitto).
\end{acknowledgements}
\bibliographystyle{aa} 
\bibliography{refs}
\end{document}